\documentclass[10pt,sigconf,letterpaper,nonacm]{acmart}

\usepackage[english]{babel}
\usepackage{times}
\usepackage{blindtext}
\usepackage{tabularx}
\usepackage{graphicx} 
\usepackage{color}
\usepackage{xspace}
\usepackage{thumbpdf}
\usepackage{listings}
\usepackage{verbatim}
\usepackage{hyperref}
\usepackage{booktabs}
\usepackage{placeins}
\usepackage{balance}
\usepackage{subfigure}
\usepackage{appendix}
\usepackage[small,compact]{titlesec}
\usepackage{url}
\usepackage{multirow}
\usepackage{bm}
\usepackage{algorithm}
\usepackage{enumitem}
\usepackage{algpseudocode}
\hypersetup{breaklinks=true}

\renewcommand\footnotetextcopyrightpermission[1]{} 
\setcopyright{none}
\acmDOI{}
\acmISBN{}
\acmPrice{}

\newcommand{\bi}{\begin{itemize}}
\newcommand{\ei}{\end{itemize}}

\newcommand\eat[1]{}
\newcommand\paragraphb[1]{\noindent{\bf #1}}
\newcommand\paragraphi[1]{\noindent\emph{#1}}
\newcommand{\allnotes}[1]{}
\renewcommand{\allnotes}[1]{\textit{#1}}

\newcommand{\oursys}{Octopus\xspace}
\newcommand{\primone}{drop-by-msg\xspace}
\newcommand{\primtwo}{drop-by-bitrate\xspace}

\newcommand{\OctoBBR}{OctoBBR\xspace}

\begin{document}

\title{Controlling Congestion via In-Network Content Adaptation}

\author{Yongzhou Chen, Ammar Tahir, Radhika Mittal \\
\emph{UIUC}}

\renewcommand{\shortauthors}{X.et al.}

\begin{abstract}

Realizing that it is inherently difficult to precisely match the sending rates at the endhost with the available capacity on dynamic cellular links, we build a system, \oursys, that sends real-time data streams over cellular networks using an imprecise controller (that errs on the side of over-estimating network capacity), and then drops appropriate packets in the cellular network buffers to match the actual capacity. We design parameterized primitives for implementing the packet dropping logic, that the applications at the endhost can configure differently to express different content adaptation policies. \oursys transport encodes the app-specified parameters in packet header fields, which the routers parse to execute the desired dropping behavior.   
Our evaluation shows how real-time applications involving standard and volumetric videos can be designed to exploit \oursys, and achieve 1.5-50$\times$ better performance than state-of-the-art schemes.

\end{abstract}

\maketitle

\section{Introduction}

Congestion control solutions predominantly rely on the endhost picking the right sending rate to match the network capacity that is estimated using network feedback. The feedback could be implicit (e.g. packet drops~\cite{cubic, new-reno} or delay~\cite{bbr, sprout, verus, vegas, timely}) or may involve more active involvement from the routers (e.g. early congestion notifications~\cite{abc, dctcp, d2tcp, dcqcn, ecn}, early drops~\cite{red, codel}, and explicit rate signalling~\cite{rcp, xcp}). We collectively refer to such schemes as feedback-based controllers. 

It is important for a feedback-based controller to precisely match the sending rate with network capacity – sending too little leads to low throughput, and sending too much leads to high queuing delays and packet drops. Inter-flow scheduling schemes (e.g. fair queuing or flow prioritization~\cite{diffserv,drr,fq,gps,pfabric,pifo}) can provide isolation across flows, but minimizing the self-inflicted queuing delay for a given flow still relies on precise feedback-based congestion control. However, as we show in \S\ref{sec:motivation}, it is not possible to precisely control sending rate using a feedback-based controller when network capacity changes at timescales that are smaller than the time taken to get feedback from the network (as is common in case of dynamic cellular links~\cite{sprout, abc, 5g-study, 5g_dataset, variegated_5g}). 

In this paper, we explore an alternative approach that sidesteps the need for a precise feedback-based controller: we allow the sender to send too much to ensure high throughput, and then adapt the transmitted content in network buffers by semantically dropping packets to match the available capacity and minimize delay. 



As a motivating usecase, we consider real-time data streams sent over cellular downlinks (e.g. for video conferencing, virtual reality, etc).~\footnote{We focus on scenarios where the cellular downlink is the bottleneck; we can handle uplink bottlenecks by similar content adaptation at the end-device.}  Real-time streams, comprising of a series of multi-packet \emph{messages} (e.g. video frames), 
can typically adapt their content (e.g. video frame rate or quality) based on available network capacity in order to achieve low message latency. The state-of-the-art real-time transmission schemes (primarily focusing on real-time video, e.g.~\cite{salsify,awstream, google-gcc, qarc}) use a feedback-based controller to estimate network bandwidth, with the application then adapting its content based on this estimate. 
However, given the inherent difficulty in accurately estimating the dynamic cellular link capacity, the feedback-based controllers either underestimate it~\cite{sprout, codel} (leading to low throughput and content quality), or overestimate it~\cite{bbr, cubic} (leading to high lag in message delivery).  

We propose sending real-time streams using an imprecise feedback-based controller (that errs on the side of overestimating network bandwidth)
and then adapting the transmitted content by dropping ``appropriate'' packets in the cellular router buffers. We build a system, \oursys, to enable such in-network content adaptation. 


So how do we go about doing in-network content adaptation?
First of all, it requires the real-time application (app) to encode its streams in a way that supports content adaptation via packet drops. As detailed in our case-studies (\S\ref{sec:casestudies}), existing stream encoding techniques (e.g. scalable video codec~\cite{h264-svc,wien2007svc, webrtc-svc} and point clouds for volumetric videos \cite{groot, vivo}) already provide such adaptability. The in-network packet dropping logic would then depend on the app's requirements and how it encodes the stream. For instance, some streams may support reducing the spatial resolution (e.g. by dropping packets corresponding to higher quality layers in a layered video stream~\cite{h264-svc, webrtc-svc}, or to higher density levels in a point cloud \cite{vivo}), whereas others may solely allow reducing temporal resolution (e.g. by dropping certain frames). 

This leads us to the following question: how do we support different packet dropping policies that may vary across applications? 
Since it is not practical for cellular routers to implement customized dropping logic specific to each app, we look for generalized dropping \emph{primitives} 
which can be configured differently by different apps to express their requirements. 
The mode of configuration we adopt involves parameters that can be specified in packet header fields.

We first considered using well-known queue management techniques for this that seemingly provide such configurability.
One option was to use priority dropping~\cite{plp, bhattacharjee1998network}, where the app marks different packets with different priorities, and the router drops lower priority packets when the buffer is full. Another option was to tag different packets with deadlines, and the router drops the packets that exceed their deadline \cite{D3}. However, it was not immediately obvious how an app could use these schemes to express their content adaptation policies that often involve complex dependencies across frames. For instance, if the app requires minimizing the latency of the latest message, how do we tune the buffer threshold for priority dropping or the packet deadlines, the optimal value for which would vary with message sizes and network bandwidth. 


So we instead design parameterized dropping primitives to more directly capture the requirements of real-time apps. Our primitives drop packets at the granularity of multi-packet messages, where the app specifies the message boundaries, and expresses its dropping policy as per-message parameters. Our primitive trigger message drops based on two natural conditions: (1) arrival of a new message, where the app can specify which messages trigger a drop in which subset of older queued-up messages using priority levels, and (2) when the link capacity falls below specified bitrate thresholds, where the app can configure different thresholds for different messages based on the (known) bitrate of the generated content. Apps can configure these primitives by setting message parameters to express different content adaptation policies, 
in a manner that is agnostic of underlying network characteristics. 

Centered on the above primitives, we build a system, \oursys, that comprises of: (i) an interface for applications to specify their message boundaries and per-message parameters to configure the dropping primitives (ii) a transport protocol that encodes the app-specified message parameters into packet headers, performs (imperfect) congestion control, and implements the parameterized dropping primitives for content adaptation in transport buffer, (iii) router buffer management scheme that implements the dropping primitives, and parses the parameters in packet header fields to enforce app-specified content adaptation policy.





We acknowledge that our system cannot be \emph{immediately} deployed, as it requires changes at both the endpoints and the cellular base-stations. However, we believe that our approach presents an interesting design point that is worth exploring, given the increasing volume and significance of real-time streaming~\cite{video_conference, metaverse} that is well-suited to in-network content adaptation, the increasing flexibility of modern network infrastructure~\cite{p4, rmt,openran}, and the scope for significant improvement in performance (as promised by our evaluation).

We summarize our key contributions below:

\begin{itemize}[leftmargin=*]

\item We present a new approach for controlling congestion that is based on adapting the transmitted content inside the network, rather than trying to perfectly match the sending rate with estimated network capacity at the endhost.


\item We design a system to realize this new approach (\S\ref{sec:design}), which is centered around parameterized primitives that can be used to express different content adaptation policies. 

\item We prototype our system (\S\ref{sec:prototype}) using UDT~\cite{udt} (a user-space transport framework) for the endhost logic, and srsRAN~\cite{srsran} (an open-source cellular platform) for the in-network logic. 

\item We evaluate our prototype using three different case-studies (\S\ref{sec:casestudies}), and across a variety of scenarios (\S\ref{sec:eval-details}), to show how our approach results in 1.5-50$\times$ better performance than state-of-the-art schemes~\cite{awstream, salsify, vivo}.

\end{itemize}

\begin{figure*}[!ht]
    \centering
    \subfigure[TCP-BBR]{\includegraphics[scale=0.21]{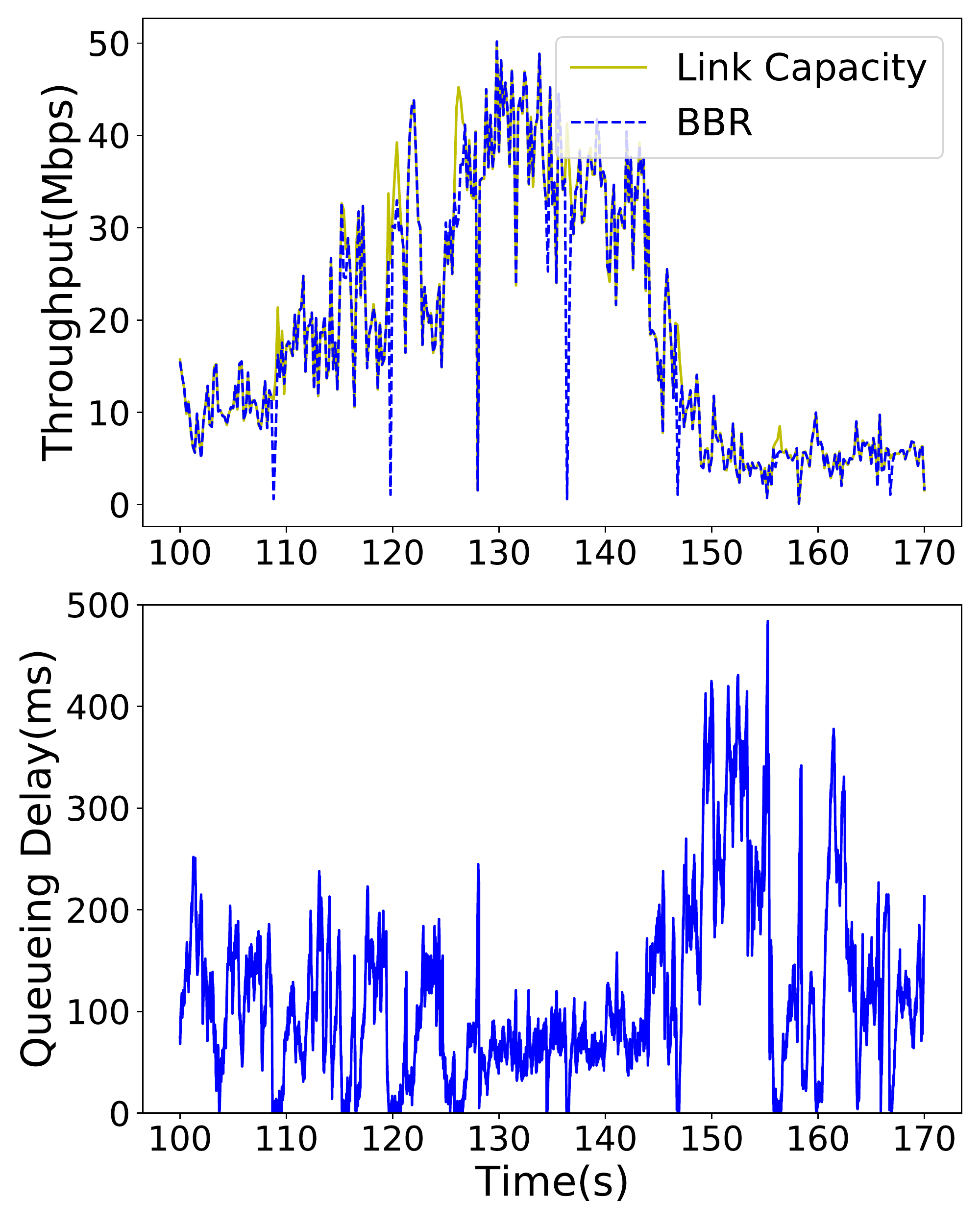}}
    \subfigure[Sprout]{\includegraphics[scale=0.21]{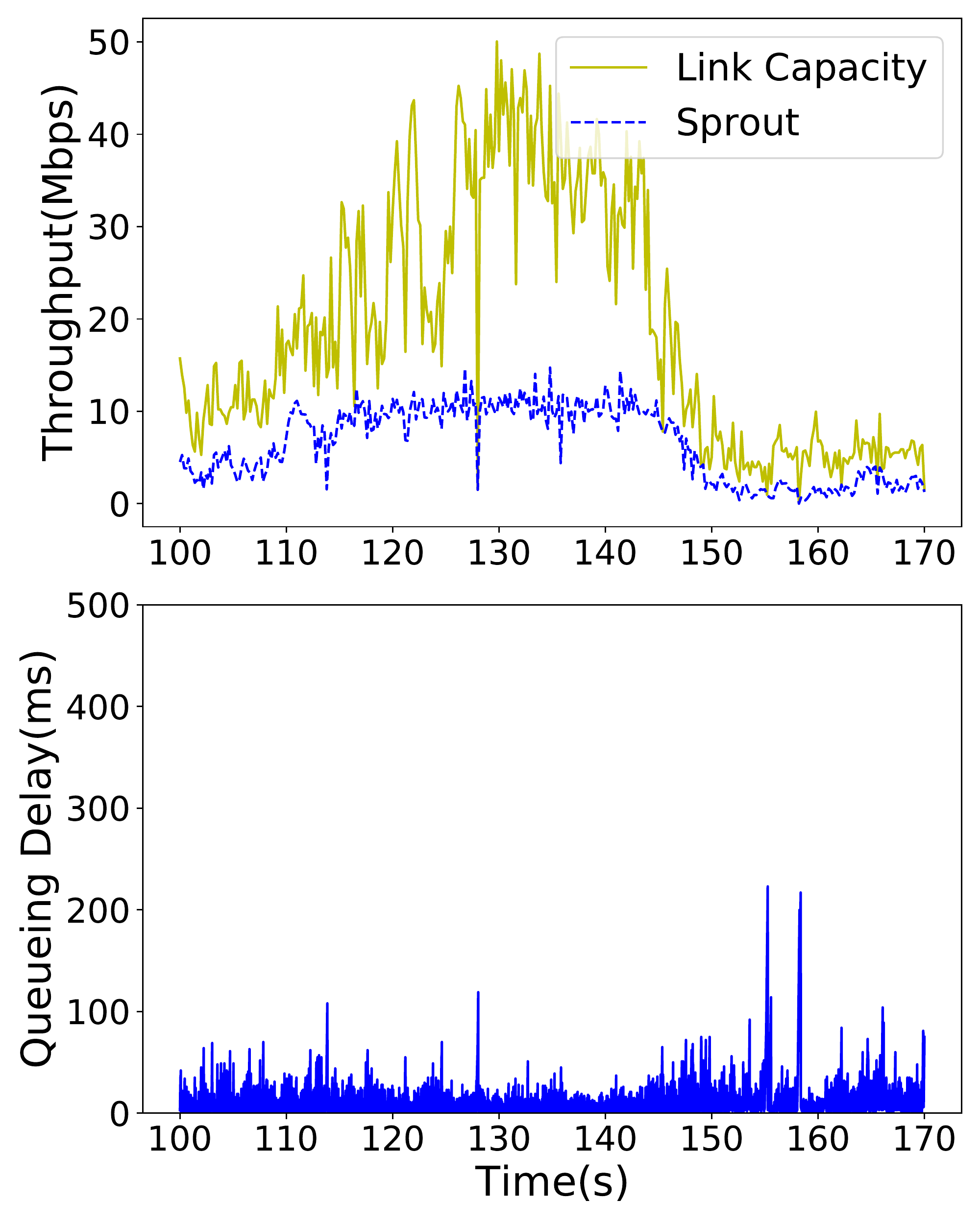}}
    \subfigure[TCP-Cubic \& Codel]{\includegraphics[scale=0.21]{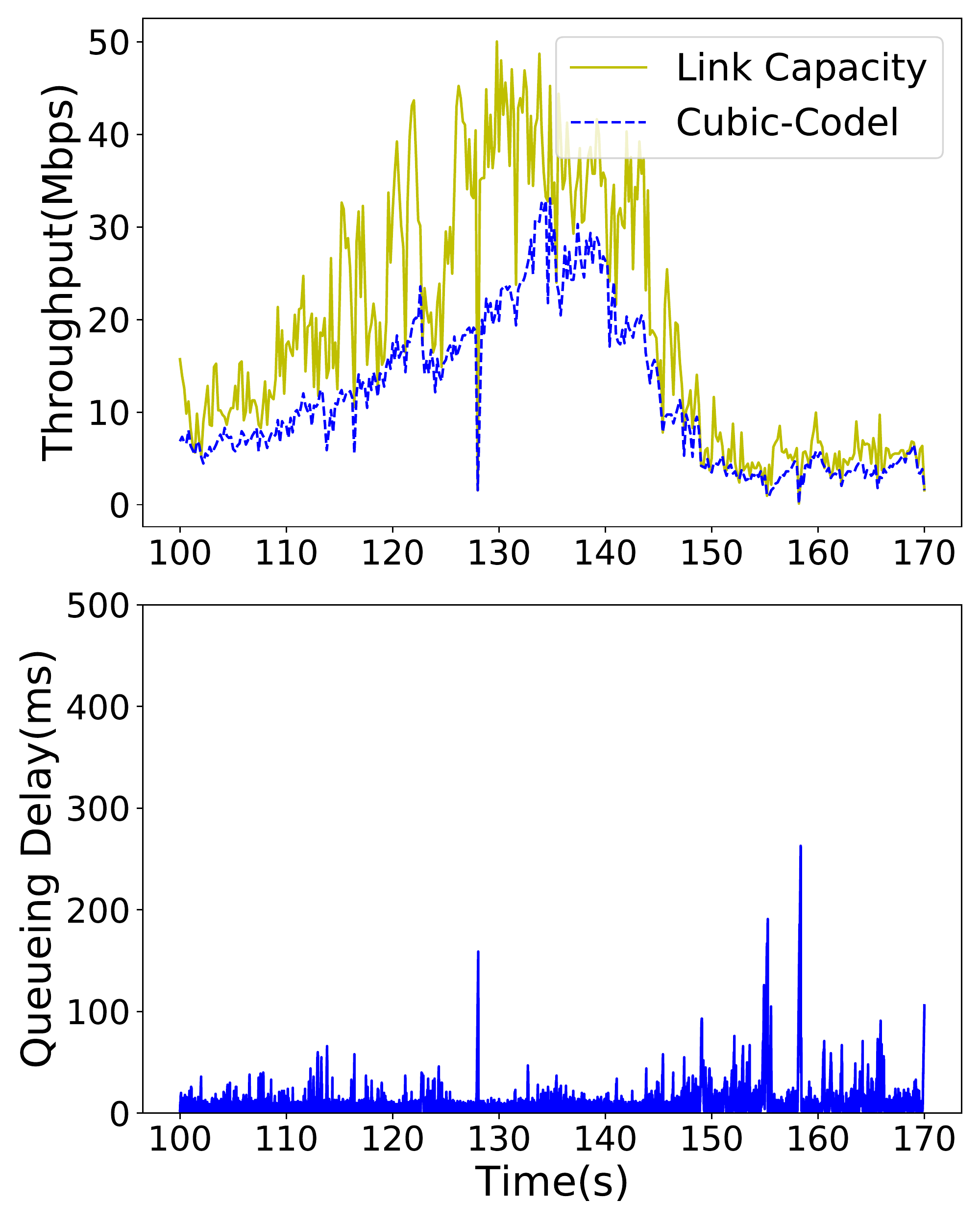}}
    \subfigure[ABC]{\includegraphics[scale=0.21]{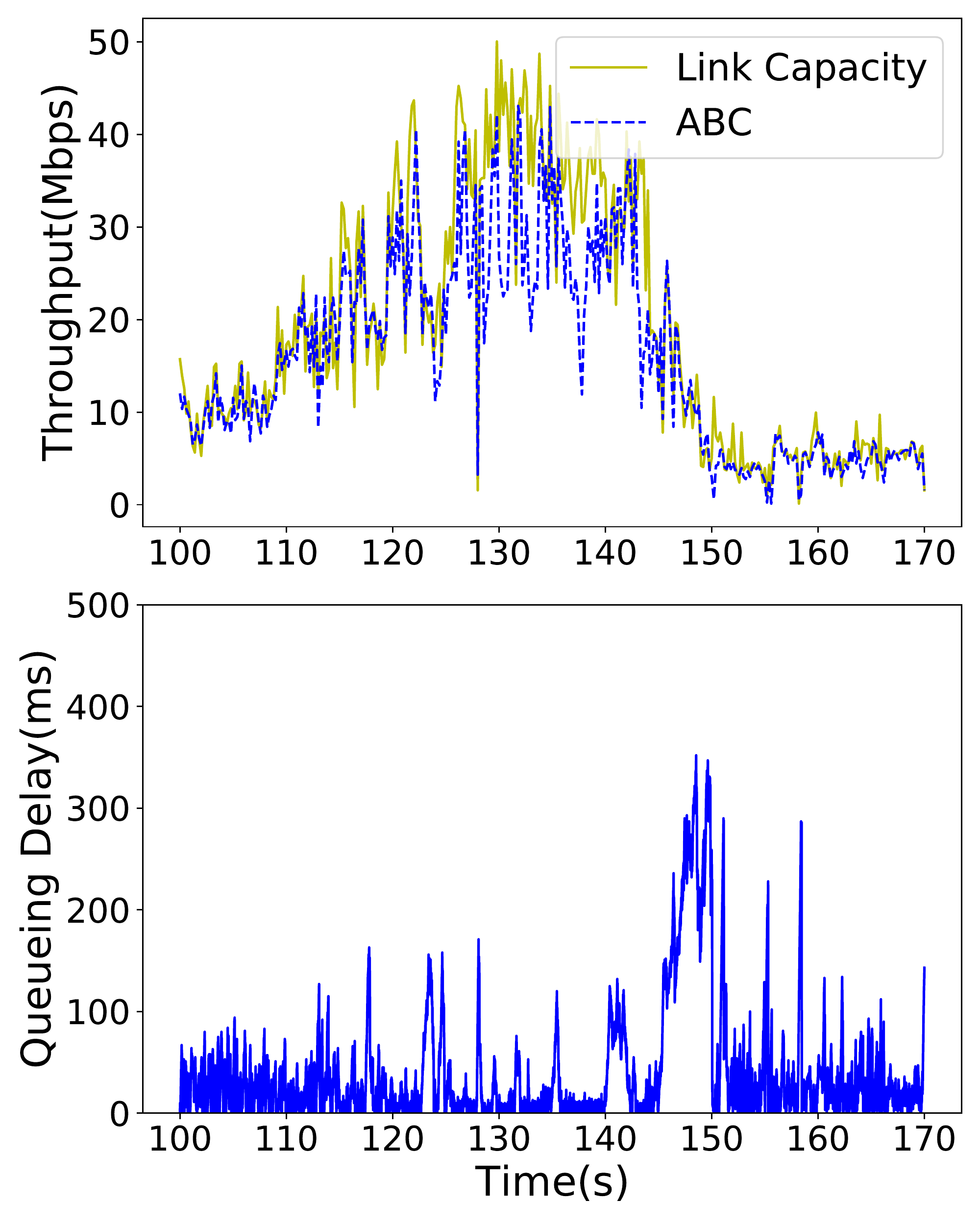}}
    \vspace{-10pt}
    \caption{Throughput (top) and queuing delay (bottom) for different protocols on a link emulating Verizon download trace with RTT 60ms. We cut-off the delay graphs at 500ms to better highlight queuing over time.} 
    \vspace{-10pt}
    \label{fig:motivation}
\end{figure*}

\section{Motivation}
\label{sec:motivation}

The cellular link from the base-station to the user device is often the bottleneck for data transmission~\cite{sprout, mowie, bursttracker}.
These links are prone to bandwidth (capacity) variations~\cite{sprout, verus, abc}, triggered by dynamic obstacles or changing directionality or distance as a device moves. 
We conducted a series of experiments to evaluate how well a feedback-based congestion controller performs on a volatile cellular network. We used Mahimahi~\cite{mahimahi} to emulate a link with 60ms base round-trip time (RTT), 375KB buffer size, and time-varying bandwidth drawn from a Verizon trace~\cite{mahimahi}. A sender attached to the emulated link sends backlogged data to a receiver, using different congestion controllers.

\paragraphb{Prior works fail to achieve both high link-utilization and low queuing.} Figure~\ref{fig:motivation}(a) shows that TCP BBR\cite{bbr} (designed for WAN) achieves high link utilization, but also results in high queuing delays, that reach up to 500 milliseconds. 
We see similar results with TCP Cubic (not shown for brevity). This led to development of congestion controllers for cellular networks that react faster to changing network bandwidth~\cite{sprout, abc, verus}. Our experiment with Sprout~\cite{sprout} revealed how it can be too cautious resulting under-utilization of link capacity, and even then it cannot avoid spikes in delay when bandwidth suddenly plunges (Figure~\ref{fig:motivation}(b)). Schemes that use a feedback-based congestion controller with AQM, such as Cubic+Codel and ABC\cite{abc}, show similar issues of under-utilization and delay spikes (Figure~\ref{fig:motivation}(c,d)).


\begin{figure}[h!]
    \centering
    \includegraphics[width=0.48\textwidth]{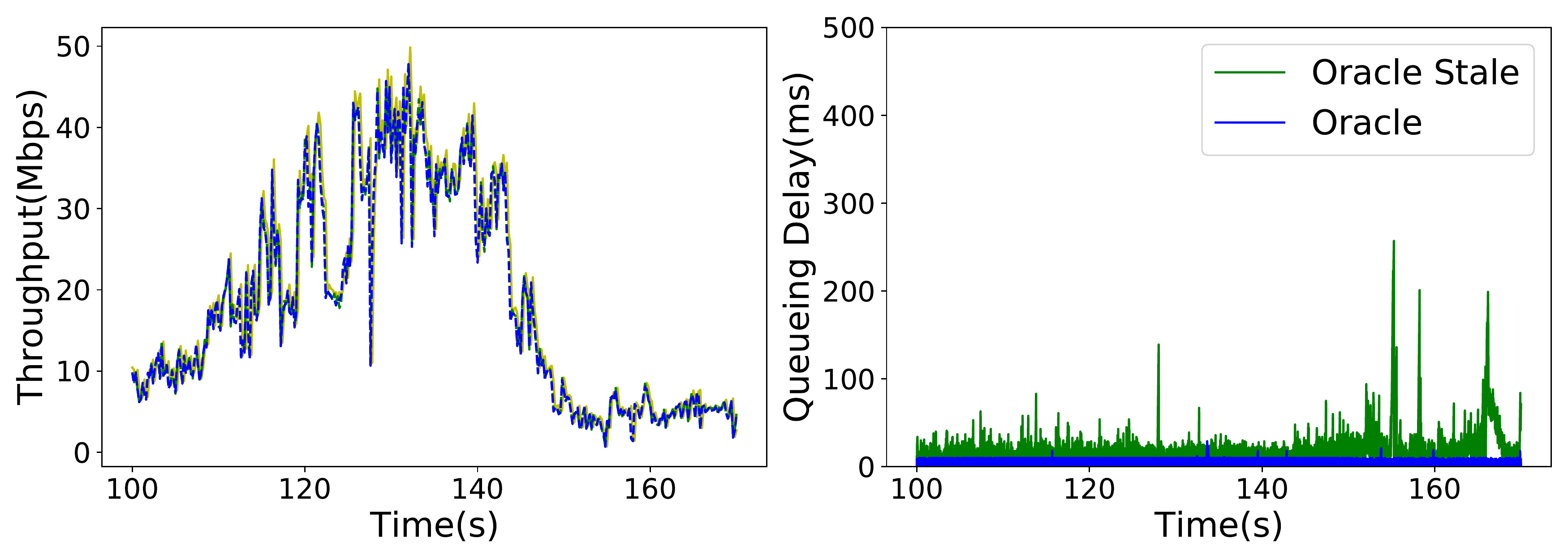}
    \vskip -0.1 in
    \caption{Throughput and queuing delay for Oracle on a link emulating Verizon download trace with RTT 60ms.}
    \vskip -0.3 in
    \label{fig:motivation-oracle}
\end{figure}

\paragraphb{We cannot practically design a perfect congestion controller!} The above were but a few samples from the vast repertoire of congestion control algorithms designed for wide-area and cellular networks~\cite{cubic, bbr, sprout, verus, abc, pcc, pcc-vivace, pbe-cc}. 
This raises the natural question of whether a different controller would have perfectly achieved high link utilization and consistently low queuing delay? Rather than taking up the insurmountable task of experimenting with all proposed congestion controllers, we instead exploited our emulation environment to do a simple exercise. We implemented an ``Oracle'' that used the knowledge of the emulated bandwidth trace to compute the number of packets that the link can serve over each period of 5ms, and sent precisely those many packets during that time period. We observed that this indeed resulted in perfect behavior (maximal link utilization with queuing delay consistently below 30ms). 

We then added a small offset, making the sending rate stale by 5ms compared to the link capacity. This staleness models real-world scenarios where an endhost can receive network feedback only after some inevitable delay.

Figure~\ref{fig:motivation-oracle} shows the resulting performance. We find that even with an offset as small as 5ms, our Oracle was unable to avoid spikes in queuing. The queue keeps accumulating packets that arrive 5ms late during a bandwidth reduction 
(thus missing their chance to be transmitted), until the next increase in bandwidth allows transmitting them.

These results show that spikes in queuing delay cannot be avoided without knowledge of (i) precisely how long a packet would take to reach the base-station, and (ii) the available bandwidth at that (future) time which may change based on external factors such as a sudden obstacles. Since it is not possible to obtain such information in practice, we do not expect any feedback-based congestion controller to behave perfectly.






\paragraphb{Our approach.}
Given the pessimistic results above, how can we meet the stringent throughput and latency requirements of networked real-time apps?  Our system, \oursys, side-steps the problem of designing a perfect feedback-based congestion controller by using a reactive approach that exploits the content adaptability of real-time streams. \oursys sends data using BBR congestion control to achieve high link utilization, and then drops appropriate packets in the cellular network buffers to match the actual available link capacity and minimize queuing delay. 
We design parameterized primitives for implementing the dropping logic, that the applications at the endhost can configure differently to express different content adaptation policies. \oursys transport encodes the app-specified parameters in packet header fields, which the routers can parse to execute the desired dropping behavior.  
We provide a detailed description of \oursys' design in \S\ref{sec:design}. 
\section{Related work}
\label{sec:related}



\paragraphb{Packet scheduling.} 
Scheduling packets \emph{across} different flows (e.g. to achieve fairness or small flow completion time)~\cite{diffserv,drr,fq,gps,pfabric,pifo} is orthogonal and complementary to our goals of ensuring optimal throughput and latency for a given real-time flow. 
%
In the context of intra-flow scheduling, recent work~\cite{bedewey2019, kadota2019} analyze the benefits of using LCFS (last-come-first-serve) for maximizing freshness at per-packet granularity. Since a real-time message is typically consumed by the receiver as soon as it is received, there is little value to delivering an older message out-of-order, after a later message in the stream has been delivered. Therefore, rather than determining the packet scheduling order, the key question we consider is  whether a message in a given real-time stream should be transmitted by the router or dropped altogether. We enforce our (more general) dropping primitives at the granularity of multi-packet messages. 

\paragraphb{In-network dropping policies.} CoDel~\cite{codel} and RED~\cite{red} proactively drop packets on the onset of congestion to send an early signal to an endhost congestion controller (\S\ref{sec:motivation} shows how this can lead to link under-utilization). \oursys, instead, drops packets to directly adapt the transmitted content based on the network conditions and app-specified requirements. 

Bhattacharjee et. al. proposed intelligent packet discard for MPEG transfers~\cite{zegura-an,bhattacharjee1998network} as a usecase of active networking. The goal was to ensure that a buffer overflow discards lower priority video frames. 
\oursys differs in its goal of minimizing latency by proactively discarding messages (before the buffer fills up) using more direct triggers in the form of new message arrivals and instantaneous link capacity. We evaluate the significance of doing so in \S\ref{sec:eval-details}.
Bhattacharjee et. al.'s proposal further required the routers to maintain app-specific logic.  
In contrast, \oursys routers are app-agnostic, and implement parameterized primitives to enforce app-specified policies.


\paragraphb{Video adapation at endhosts.} Most current schemes for real-time video streaming adapt quality or frame rate at the endhosts, relying on a congestion control algorithm to estimate network capacity~\cite{salsify, awstream, google-gcc, qarc, rebera, zhu2012}; this includes schemes that use SVC~\cite{wien2007svc, schierl2007svc} but drop higher quality layers at the endhost. Inaccuracies in bandwidth estimation at the endhost may result in sub-optimal throughput or higher latencies (\S\ref{sec:motivation}).

\section{\oursys Design}
\label{sec:design}


\begin{figure}
    \centering
    \includegraphics[width=0.45\textwidth]{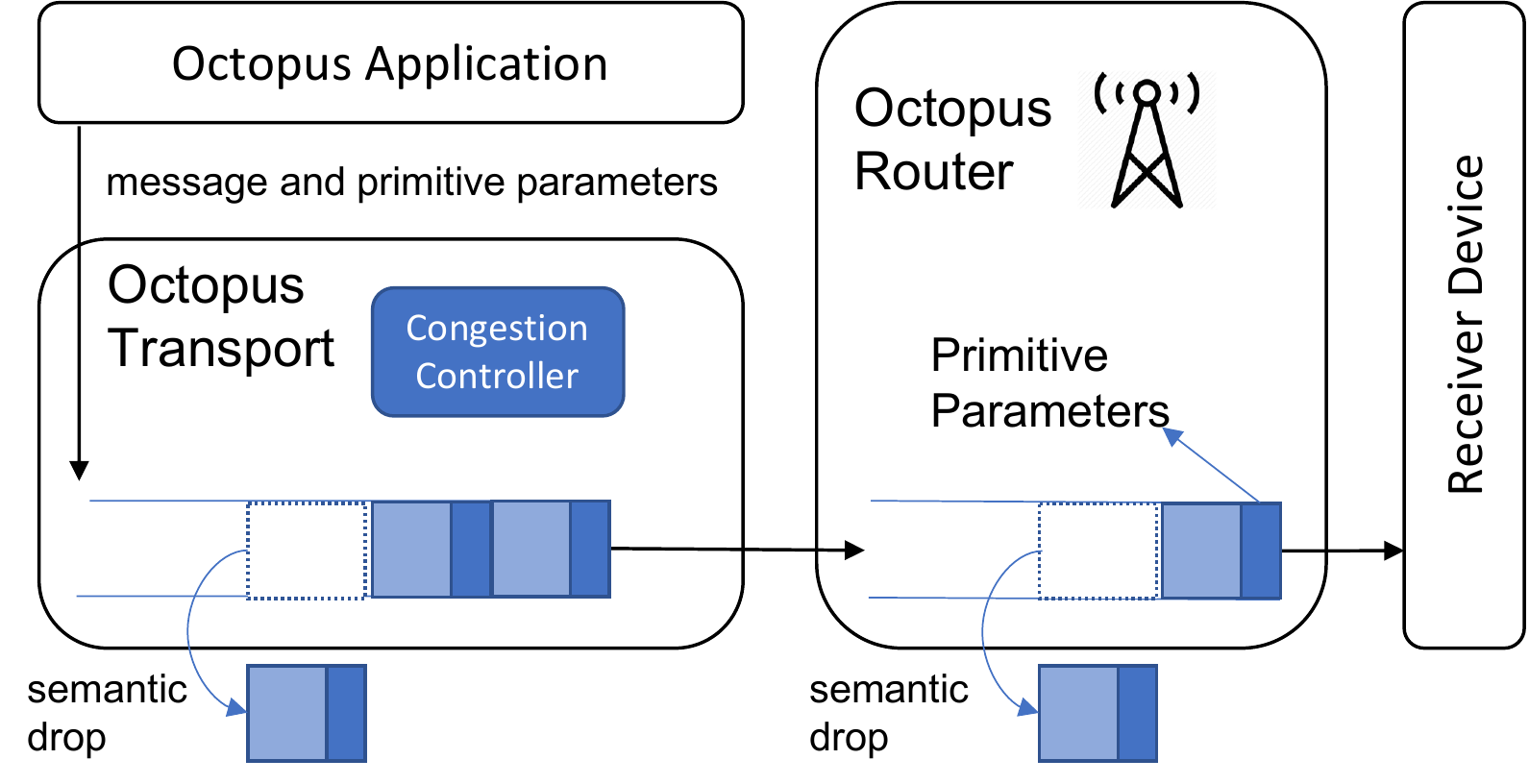}
    \caption{Octopus Framework Overview}
    \label{fig:overview}
\end{figure}

As depicted in Figure\ref{fig:overview}, \oursys is a cross-layer system consisting of the following elements:

\paragraphb{\oursys-aware Application.} 
An app specifies its desired dropping policy (as per its requirements and stream encoding format) to the underlying \oursys transport in the form of parameters for \oursys' \emph{dropping primitives} (detailed in \S\ref{sec:primitives}). The app specifies these parameters for individual \emph{messages} (where a message may contain one or more packets, and is the unit of packet drops in \oursys). 

This seemingly increases the burden on app development. Nonetheless, app-layer frameworks that support mechanisms to encode real-time streams and adapt their content based on estimated bandwidth availability already exist (e.g. WebRTC for real-time videos\cite{webrtc}). We envision that rather than changing individual apps, one would extend such frameworks to exploit \oursys. 

In order to fully exploit \oursys, a real-time stream must be able to tolerate in-network packet drops. For instance, in a single-layered video codec that uses every single frame as a reference to encode the next frame (as in H265~\cite{h265} and VP9~\cite{vp9}), a drop in one frame will disable decoding all subsequent frames. We show how existing multi-layer extensions of such codecs (e.g. SVC~\cite{h264-svc, vp9-spatial}) can be effectively used with \oursys in \S\ref{sec:casestudies}.

\paragraphb{\oursys Transport.} The underlying transport (detailed in \S\ref{sec:transport}) encodes the dropping policies (that the app specifies in the form of per-message parameters) in individual packet headers. It paces out the data packets sent into the network using BBR's congestion control logic. This ensures that the data transfer rate is limited to the peak capacity of the bottleneck cellular link, and that the \oursys traffic competes fairly with the rest of the wide-area traffic. If the pacing rate is slower than the rate at which an app generates data, messages get queued up in the transport buffer. \oursys implements its message dropping primitives at the transport buffer (as per the app-specified parameters). This direct and timely content adaptation in the transport buffer mitigates the need for additional app-layer content adaptation strategies that are based on feedback from the transport.~\footnote{We provide a direct comparison of \oursys' endpoint content adaptation (without the in-network logic) with other endpoint-based solutions in \S\ref{sec:octobbr}.} It acts as a first-level of content adaption at the endpoint itself, and is sufficient if the network bandwidth is stable and matches the pacing rate.

\paragraphb{\oursys Router.} Given the inherent inaccuracy in estimating volatile link bandwidth, and the relatively aggressive congestion controller (BBR) used by \oursys, the pacing rate at an endpoint might be higher than what the cellular link can support. The \oursys router logic (\S\ref{sec:router}) kicks in to adapt stream content and minimize in-network delays in such cases. It implements the \oursys dropping primitives, parses the packet headers to read app-specified parameters, and enforces the desired dropping behaviour.




\begin{figure}[t!]
    \centering
    \subfigure[\normalsize{\primone}]{\includegraphics[width=0.2\textwidth]{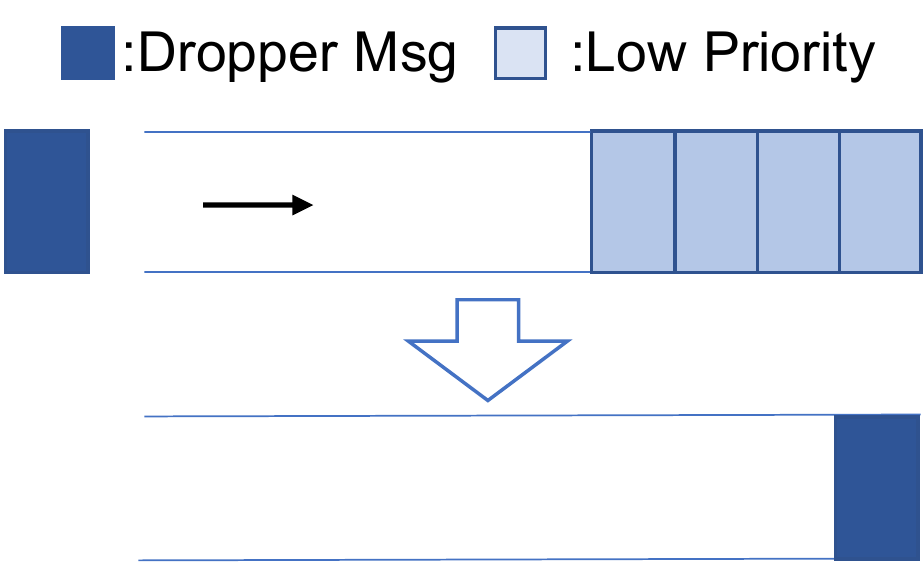}} \quad
    \subfigure[\normalsize{\primtwo}]{\includegraphics[width=0.23\textwidth]{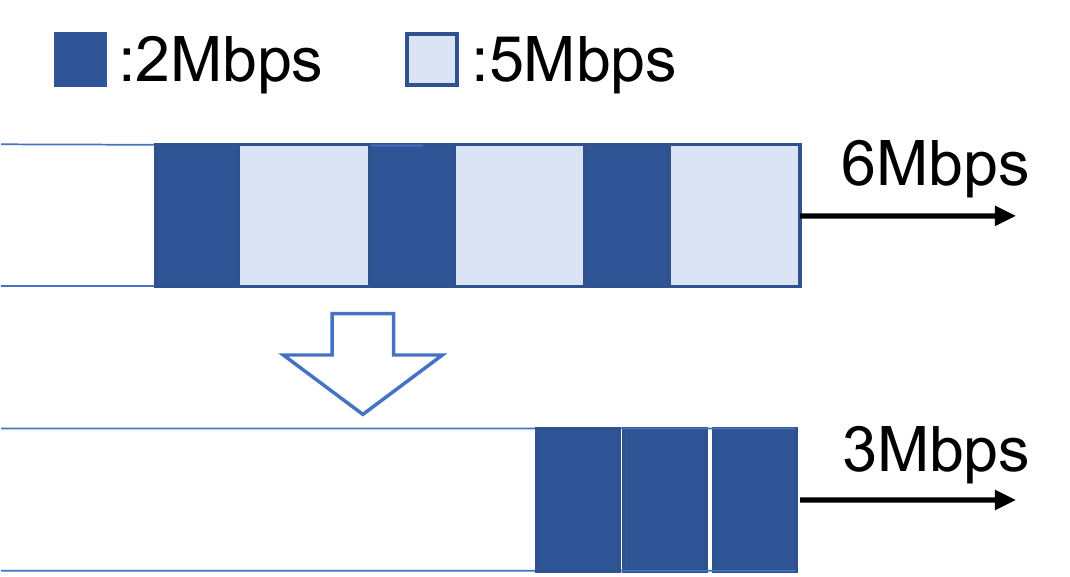}}
    \caption{Two dropping conditions for \oursys}
    \label{fig:drop-primitive}
    \vspace{-10pt}
\end{figure}

\subsection{Dropping Primitives}
\label{sec:primitives}

A real-time data stream consists of a series of temporal data units, each corresponding to a point in time. Generalizing the terminology used for video streams, we refer to each such temporal unit as a frame. Each frame has a base layer at a specific quality level, and may optionally have additional quality-enhancing layers.
If the link bandwidth is larger than the incoming rate of the stream, the link can sustain the stream and no frames get queued up. If not, multiple frames get queued up at the link buffer, and we need to adapt the buffered content by dropping packets in order to sustain the stream. There are two ways in which the content of real-time streams can be adapted -- (i) reducing the temporal resolution (or the frame rate) by dropping entire frames, and, (ii) if supported, reducing the spatial resolution (or data quality) by dropping one or more quality enhancing layers within a frame. Such content adaptation must further take the stream encoding (or the frame dependency structure) into account -- if a ``reference'' frame/layer is dropped, subsequent frames/layers that semantically depend on it cannot get decoded. 

We define a \emph{message} as the atomic granularity at which \oursys routers drop data packets in a  stream. A message would correspond to an entire frame if content is adapted solely via reducing the temporal resolution. In streams that also support adapting spatial resolution, a message would correspond to a quality layer in each frame. An app identifies its message boundaries, and specifies its dropping policies (as per its requirements and stream encoding format) in terms of per-message parameters supported by \oursys' dropping primitives. 
We now detail these primitives.




\paragraphb{Dropping Condition 1: Newer Message Arrival.}
In our first primitive, the arrival of a newer (more important) message triggers dropping (a subset of) staler queued-up messages. 
\oursys tags each message with app-specified (i) priority value (\emph{msg\_priority}), where a higher value implies lower priority, and (ii) a \emph{drop\_flag} with an associated \emph{priority\_threshold} value. When a new message is enqueued at the buffer, if its drop\_flag is set, it triggers dropping all previously queued up messages in the same stream which have  msg\_priority $\ge$ priority\_threshold. We refer to the messages with the drop\_flag set as \emph{dropper messages}, and refer to this primitive as \emph{\primone} primitive. 


An app can configure the message boundaries and the primitive's parameters to specify which messages' arrival will drop which subset of queued-up messages. For streams with independent frames (e.g. a stream of images), simply setting the drop\_flag in each message (frame) will maximize freshness. If a stream requires fixed temporal resolution but allows tuning spatial resolution, the app can configure the message boundaries and parameters to only drop the queued-up quality-enhancement layers upon the arrival of a new frame. 
Or, based on the stream encoding, an app can configure the parameters to avoid dropping a reference frame upon the arrival of a new frame dependent on it. 

In general, a queue build-up of two or more frames is a good indicator of the inability of the link to sustain the incoming stream. The \primone primitive is designed to immediately react to such queues in order to minimize queuing delay. 
However, small transient queues of multiple frames may build-up when there is a mismatch between the average and the instantaneous rate of the incoming stream due to differences in frame sizes. Our second primitive is designed to provide increased tolerance towards such transient queues.  




\vspace{10pt}
\paragraphb{Dropping Condition 2: Bandwidth Lower than Data Bitrate.} In certain stream encodings, the frame sizes may differ significantly. For example, in a layered video codec, the size of first reference frame in each ``group of pictures'' (on which subsequent frames in the group depend) can be 4X larger than the non-reference frames. Even if the link bandwidth is sufficient to sustain the average bitrate across several frames, a transient queue might build-up while serving the larger (reference) frames.
If the dropper messages in \primone primitive are spaced very closely, the quality-enhancement layers in the reference frame might get dropped in the transient queue causing the subsequent dependent frames to be decoded at the lowest quality level (even though the link bandwidth was sufficient to sustain the average bitrate of high quality layers). On the other hand, if the dropper messages are spaced too far apart (or configured to avoid dropping reference layers), and the link bandwidth is indeed smaller than the average bitrate of higher-quality layers, it could happen that by the time a suitable dropper message arrives, the higher-quality layers (that should have been dropped) have already left the queue after having contributed to a large queuing delay for the remaining frames. 

To handle such scenarios, we additionally need a dropping condition that is directly based on the available link bandwidth, and does not rely on subsequent dropper messages. Our second primitive provides this.

An app can tag each message with a \emph{bitrate\_threshold}.~\footnote{It can use its knowledge of the stream encoding and data rate for this.} \oursys drops a message if the stream is currently being served at a bandwidth $BW$ that is less than its bitrate\_threshold.
We refer to this primitive as \emph{\primtwo} primitive. It allows an app to specify different bitrate\_thresholds for different spatial layers in the stream, thus enabling the \oursys buffer to directly determine whether a spatial layer should be dropped or transmitted based on the available link capacity.

\paragraphb{Summary.} \oursys supports two conditions to drop messages. The combination of the two primitives provides an expressive mechanism for real-time apps to specify different content adaptation policies. We exemplify their usage in \S\ref{sec:casestudies}. 

\subsection{Transport Design}
\label{sec:transport}

\paragraphb{API.} An app conveys its dropping policies via \oursys' transport interface. In particular, the app conveys its atomic message boundaries to the transport, and for each message, specifies its stream\_id (which explicitly identifies the stream that the message belongs to), msg\_priority, and the dropping parameters (that include the drop\_flag, the associated priority\_threshold, and the bitrate\_threshold). 

By default, \oursys transport sets all dropping parameters in a message to zero, if not explicitly configured by the app, that disables any content adaptation. 


\paragraphb{Encoding dropping policies.}
\oursys transport tracks per-stream message sequence space. Upon receiving a new message from the application, it increments the corresponding msg\_id counter, packetizes the data, and encodes the msg\_id, priority, and parameters in designated packet header fields. The header fields additionally carry information about whether the packet is the first, last, or only packet of the message.

\paragraphb{Transport buffer management.} Similar to TCP, \oursys transport enqueues the packets in a send buffer, until they can be sent out to the lower layers. As mentioned earlier, it enforces the same message dropping policies in this transport buffer as the one enforced by the router buffer (described in more details in \S\ref{sec:router}).~\footnote{\oursys design can be naturally extended to scenarios where the bottleneck is at the upload link on the user device (and not at the download link on the cellular base-station). The endhost could use a back-pressure based mechanism (similar to TCP small queues~\cite{tcp-small-queues}) to restrict the transport from sending more data when the (small) NIC buffer is full. The \oursys logic implemented at the transport buffer would then appropriately drop messages. We leave a detailed implementation of this to future work.} It uses the bandwidth estimated by the congestion control algorithm as the current available bandwidth for the \primtwo primitive. 

\paragraphb{Loss handling and message delivery.} 
\oursys transport is unreliable (future extensions can support partial reliability). The receiver acknowledges received data (required for congestion control), although no packets are retransmitted. As soon as a message is completely received, the \oursys receive-side transport delivers it to the application, irrespective of whether prior messages have been delivered.

\paragraphb{Congestion control.}  
\oursys requires congestion control to ensure that the data transfer rate is limited to the peak capacity of the bottleneck cellular link, so as to not overwhelm the rest of the network. Moreover, there might be scenarios where the bottleneck lies elsewhere in the network (e.g. at a legacy switch that does not support \oursys), requiring \oursys to compete fairly with the cross-traffic. 
We incorporate BBR's congestion control logic in \oursys transport. BBR, by design, tracks the maximum packet bandwidth and the minimum RTT over a configurable period of time, and uses this to compute the delivery rate and the congestion window. This enables BBR to continue sending at the peak cellular capacity, without getting perturbed by transient dips in available bandwidth. As shown in \S\ref{sec:motivation}, this does come at the cost of high queueing delays which we handle through our in-network dropping policies.

\subsection{Router Design}
\label{sec:router}

An \oursys router at a cellular base-station enforces the dropping primitives, as per app-specified policies encoded in packet headers.
Cellular routers already maintain separate queues for individual users~\cite{sprout, downlink-schedule, srsran}. We further assume that a user's real-time traffic is isolated from their non-realtime traffic (standard mechanisms for doing so already exist in cellular networks\cite{downlink-schedule, lte-qos}).

\begin{algorithm}[t]
\caption{\oursys' packet dropping logic}
\begin{algorithmic}
\State \textbf{variable} dropper\_msgs\_, msg\_in\_drop\_
\Procedure{Enqueue}{packet}
\State sid $\gets$ packet.streamID()
\If {packet.hasDropFlag() \textbf{and} packet.isTail()}
    \State threshold $\gets$ packet.priorityThreshold()
    \State dropper\_msgs\_[sid][threshold] $\gets$ packet.msgID()
\EndIf
\State buffer\_.push(packet)
\EndProcedure

\Procedure{Dequeue}{void}
\State packet $\gets$ buffer\_.pop()
\State msgid $\gets$ packet.msgID()
\State sid $\gets$ packet.streamID()

\If {packet.isHead()}
    \State prio $\gets$ packet.priority()
    \State latest\_dropper $\gets$ max(dropper\_msgs\_[sid][0],
    \State \hspace*{2em} ..., dropper\_msgs\_[sid][prio])
    \State isdrop $\gets$ msgid $<$ latest\_dropper
    \State \hspace*{2em} \textbf{or} packet.bitrateThreshold() $>$           BW[sid]
    \If {isdrop}
    \State msg\_in\_drop\_[sid] $\gets$ msgid
    \EndIf
\EndIf
\If {msg\_in\_drop\_[sid] $=$ msgid}
    \State \Return Drop(packet)
\EndIf
\State \Return packet

\EndProcedure
\end{algorithmic}
\label{algo1}
\end{algorithm}

\oursys requires the router to track the available bandwidth $BW$ for each (per-user) queue. This is often directly available in cellular links~\cite{lte-bandwidth, abc}. It can also be measured at the router by tracking the rate at which packets are dequeued and transmitted, as we do in our experiments (\S\ref{sec:prototype}).  If a given user downloads multiple real-time streams, the router computes the max-min fair rate $BW_i$ for each stream $i$ from the observed per-stream arrival rate and the overall rate $BW$ at which the user's queue is served.  


Algorithm\ref{algo1} shows the pseudo-code for the packet dropping logic. For each stream, the router maintains a table indexed by the priority\_threshold that records the msg\_id of the latest dropper message corresponding to that threshold. It updates this table when enqueuing the tail packet of a dropper message. When the head packet of a message from stream $i$ is dequeued, the router marks this message to be dropped if: (i) its bitrate\_threshold is higher than $BW_i$, or (ii) its msg\_priority is greater than or equal to the priority\_threshold of all dropper message in the queue that belong to the same stream. If a message is marked for drop, all the following packets belonging to it are dropped during dequeuing. To avoid starvation, the router does not drop a message that is in transmission (i.e. one or more of whose packets have been transmitted). 


Notice that \oursys requires maintaining very modest amount of per-stream state in the routers, which remains constant with number of packets or messages. It only grows with the number of priority threshold levels in a stream -- we expect this number to be small for most usecases (\S\ref{sec:casestudies}), and it is restricted to 8 in our prototype. 

\section{Prototype Implementation}
\label{sec:prototype}

\paragraphb{\oursys Transport.} We implement \oursys transport in 3000 LoC by extending the UDT framework, a user-space transport protocol over UDP~\cite{udt}.
This includes implementing BBR's congestion control logic. 
Our prototype uses UDT's app-layer header to encode message properties and dropping policies, which take 12 bytes. An actual deployment can instead make use of IPv4 options or IPv6 extension fields to encode \oursys parameters.

\begin{figure}[t!]
\centering
\includegraphics[width=0.45\textwidth]{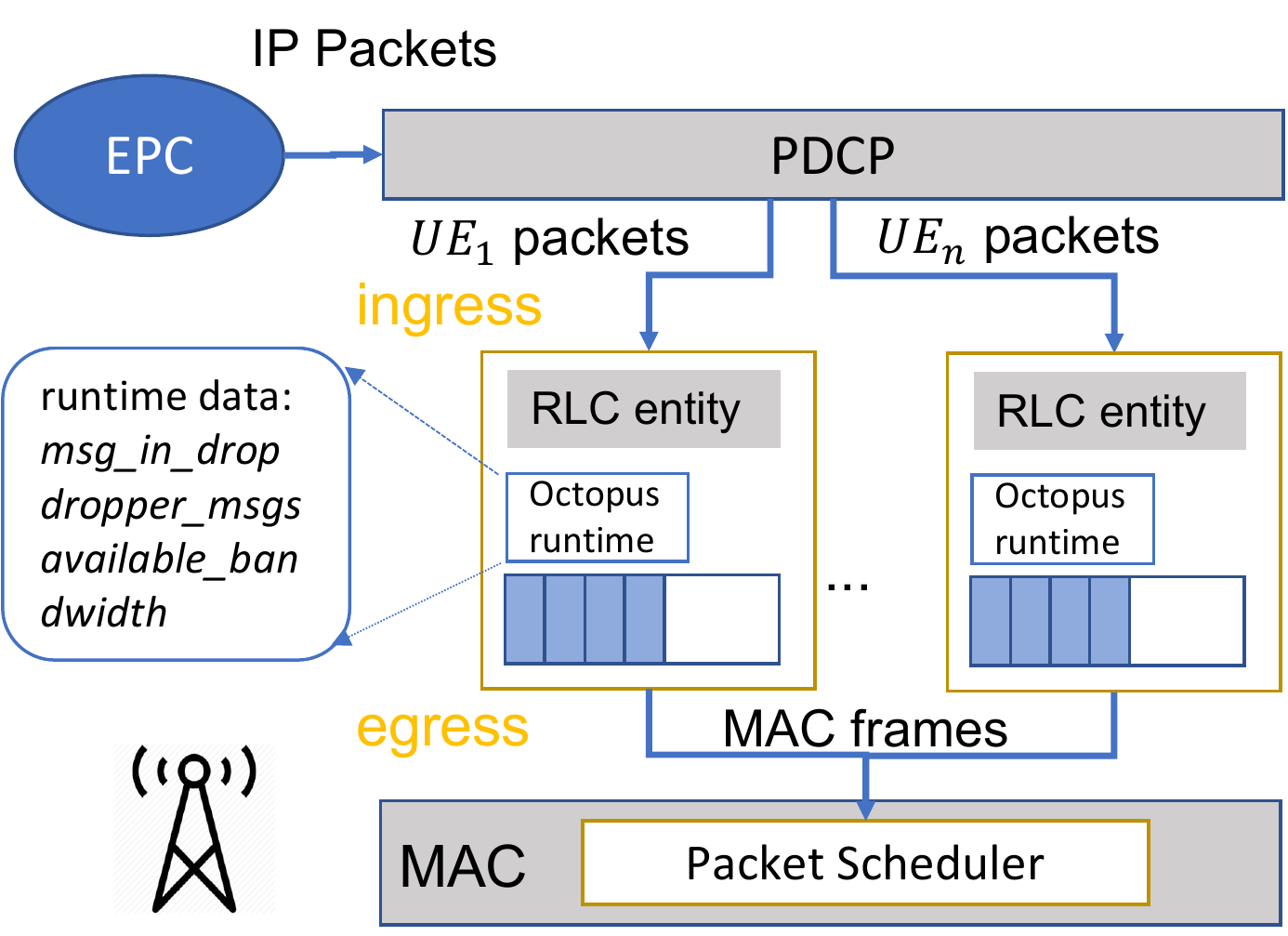}
\vspace{-10pt}
\caption{eNB/gNB protocol stack with Octopus runtime}
\vspace{-10pt}
\label{fig:eNodeB}
\end{figure}

\paragraphb{\oursys Router.} We implement Octopus' in-network logic in srsRAN\cite{srsran, srsran-code}, an open-source software cellular platform for 5G and LTE radio access network (RAN). Our changes to srsRAN fit within 420 LoC. 

The network protocol stacks of the 5G/LTE base station in srsRAN are shown in Figure~\ref{fig:eNodeB}. Specifically, the RLC(radio link control) layer instantiates an entity to manage an isolated logical queue for each connected user. We implement the Octopus' dropping primitives and maintain runtime data in the RLC entity. In the ingress stage, the RLC entity parses the app-layer headers of incoming packets, and accordingly updates the table of latest dropper messages. In the egress stage, Octopus determines whether the current message is to be dropped based on the dropping conditions, and drops packets belonging to it in that case. For the drop-by-bitrate primitive, we track the bandwidth availability by computing the dequeueing rate over a sliding time window of 50ms(discounting the idle periods when the queue is empty).

We deploy the srsRAN platform on a server with eight Intel Xeon E5 cores.
The software-based srsRAN platform can support a data-rate of up to 75Mbps -- this remained unaffected by adding \oursys logic. \oursys's dropping logic is light-weight, and can be easily supported in other hardware-based RAN platforms (we back this claim by discussing the feasibility of implementing \oursys on P4 in Appendix~\ref{appendix-p4}). 

\section{Case Studies}
\label{sec:casestudies} 

\begin{figure*}[t!]
    \centering
    \subfigure[\normalsize{ATT}]{\includegraphics[width=0.32\textwidth]{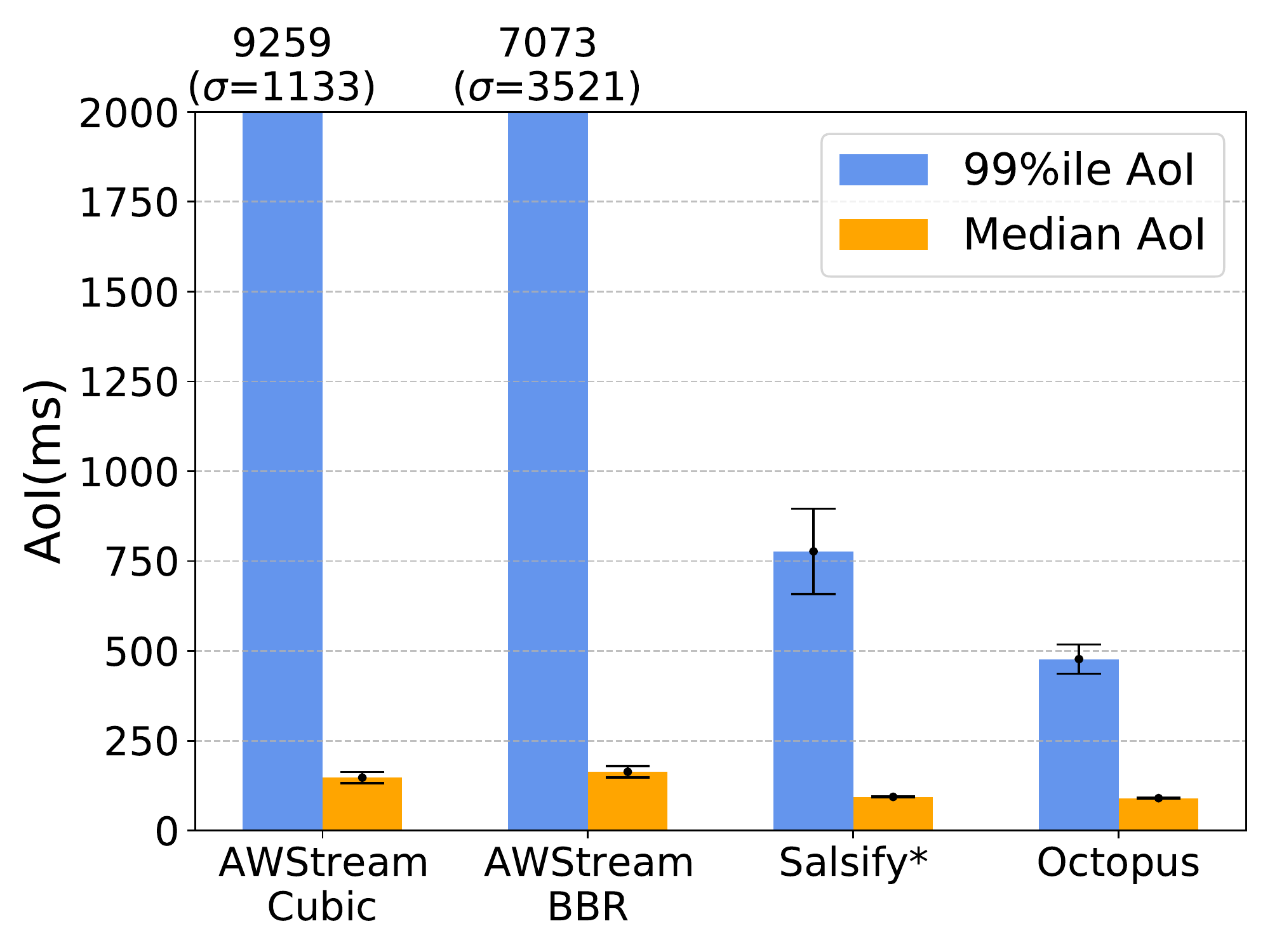}}\quad
    \subfigure[\normalsize{TMobile}]{\includegraphics[width=0.32\textwidth]{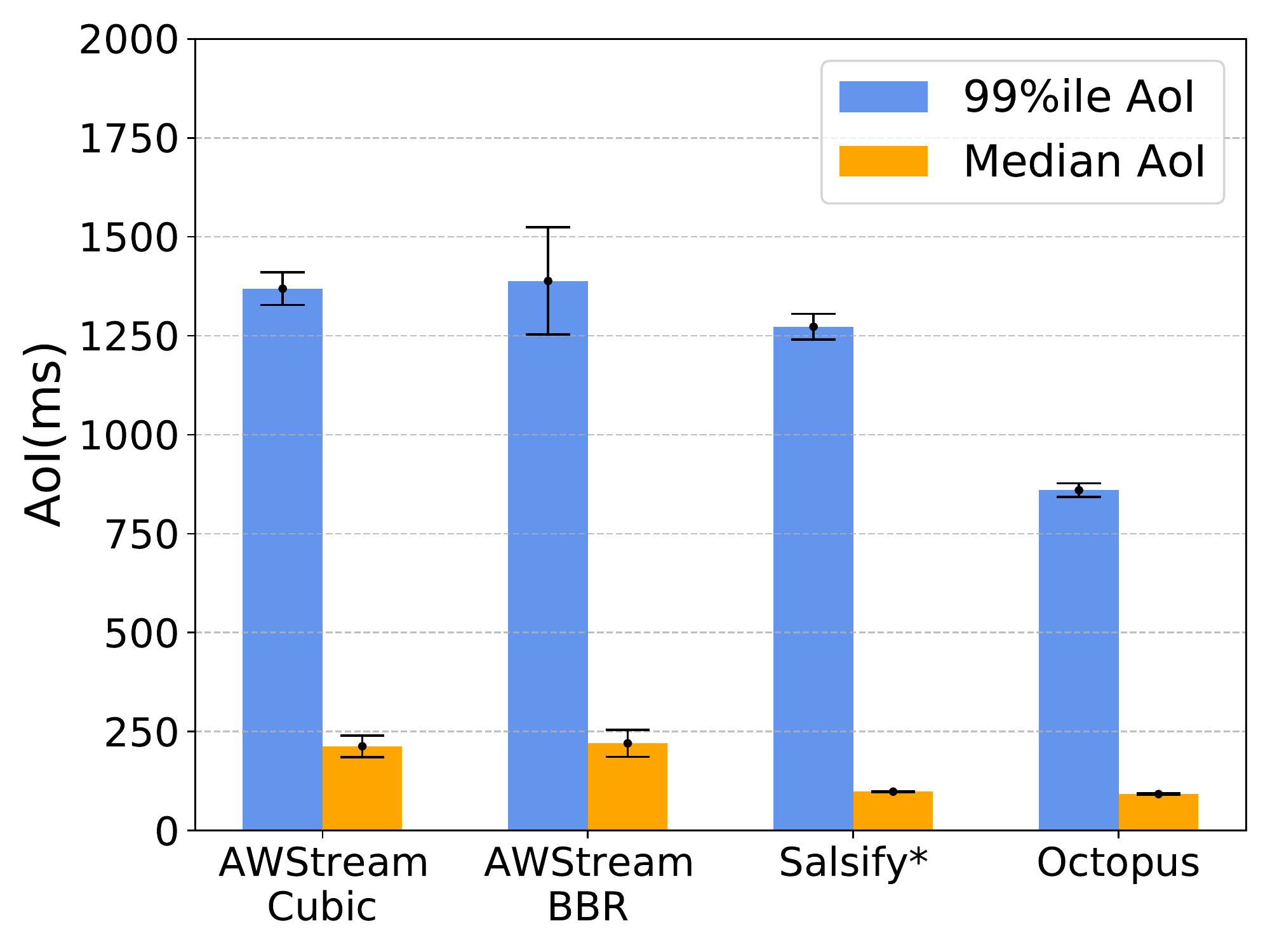}}\quad
    \subfigure[\normalsize{Verizon}]{\includegraphics[width=0.32\textwidth]{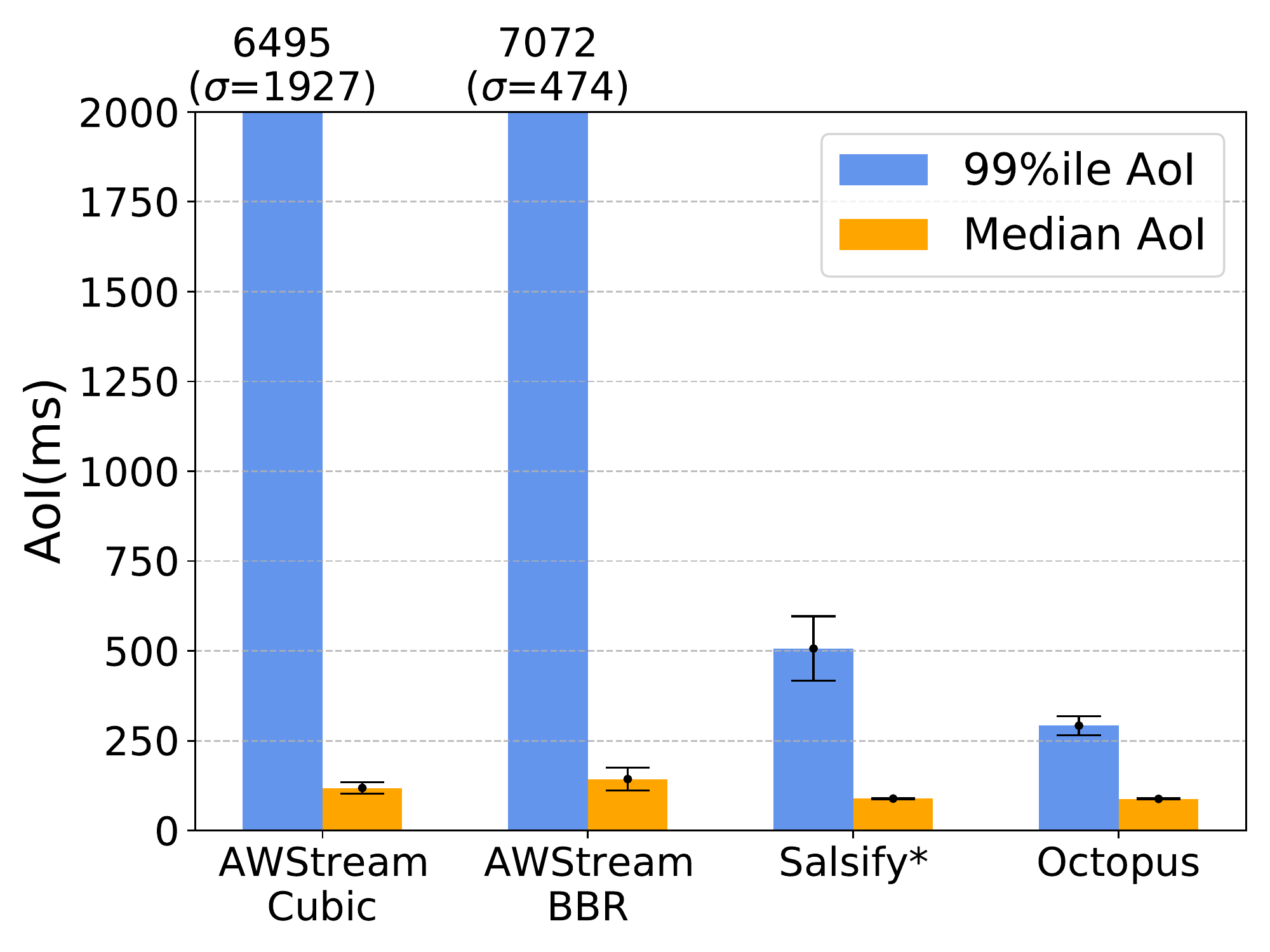}}
    \vskip -0.1 in
    \caption{Median and 99\%ile AoI in different LTE download network traces with RTT 60ms (averaged over 5 videos, with error bars showing standard deviation).}
    \label{fig:eval-usecase1}
\end{figure*}

We evaluate \oursys using three case studies involving real-time video streaming with frame rate adaptation (\S\ref{sec:casestudy1}) and quality adaption (\S\ref{sec:casestudy2}), and live volumetric video streaming (\S\ref{sec:casestudy3}). We focus on real-time 2D/3D video streaming in our case studies due to their relative popularity across apps (in conferencing, gaming, VR, surveillance, robotics, etc) and due to the existence of comparative baselines.
One can design similar policies to exploit \oursys for other forms of real-time streams. In this section, we present our basic results, comparing \oursys policies with state-of-the-art transmission schemes\cite{awstream, salsify,vivo}, and present a more in-depth evaluation in \S\ref{sec:eval-details}.

\paragraphb{Experiment Testbed.}
Our testbed involves a sender and a receiver node (both running \oursys transport on UDT) communicating via the srsRAN platform. We emulate cellular link bandwidth and delay on the downlink from the base station to the receiver.~\footnote{For this, we use the cellular downlink traces~\cite{sprout} as input to set the maximum MAC frame size in srsRAN every TTI.}
For the first two case studies, we experiment with bandwidth traces from three different LTE cellular providers (Verizon, T-Mobile, and ATT)\cite{sprout}. 
The video sources are five different videos taken from MOT17 and MOT20 datasets~\cite{mot17, mot20}. 

For the third case study involving volumetric videos, we experiment with two different 5G cellular download traces~\cite{5g_dataset}. To support higher 5G data-rates in this case-study, we replace our srsRAN platform with Mahimahi network emulator~\cite{mahimahi}, and implement \oursys' router logic in Mahimahi. 
The volumetric video sources are three videos in point cloud format taken from CMU panoptic dataset~\cite{panoptic}. 

We experiment with three different RTTs (20ms, 60ms, and 120ms) for each case-study. For brevity, we only present results with 60ms RTT (we see similar trends at other RTTs, and present results with RTT 120ms in \S\ref{sec:eval-details} and appendix).

\begin{figure}[]
\centering
\includegraphics[width=0.48\textwidth]{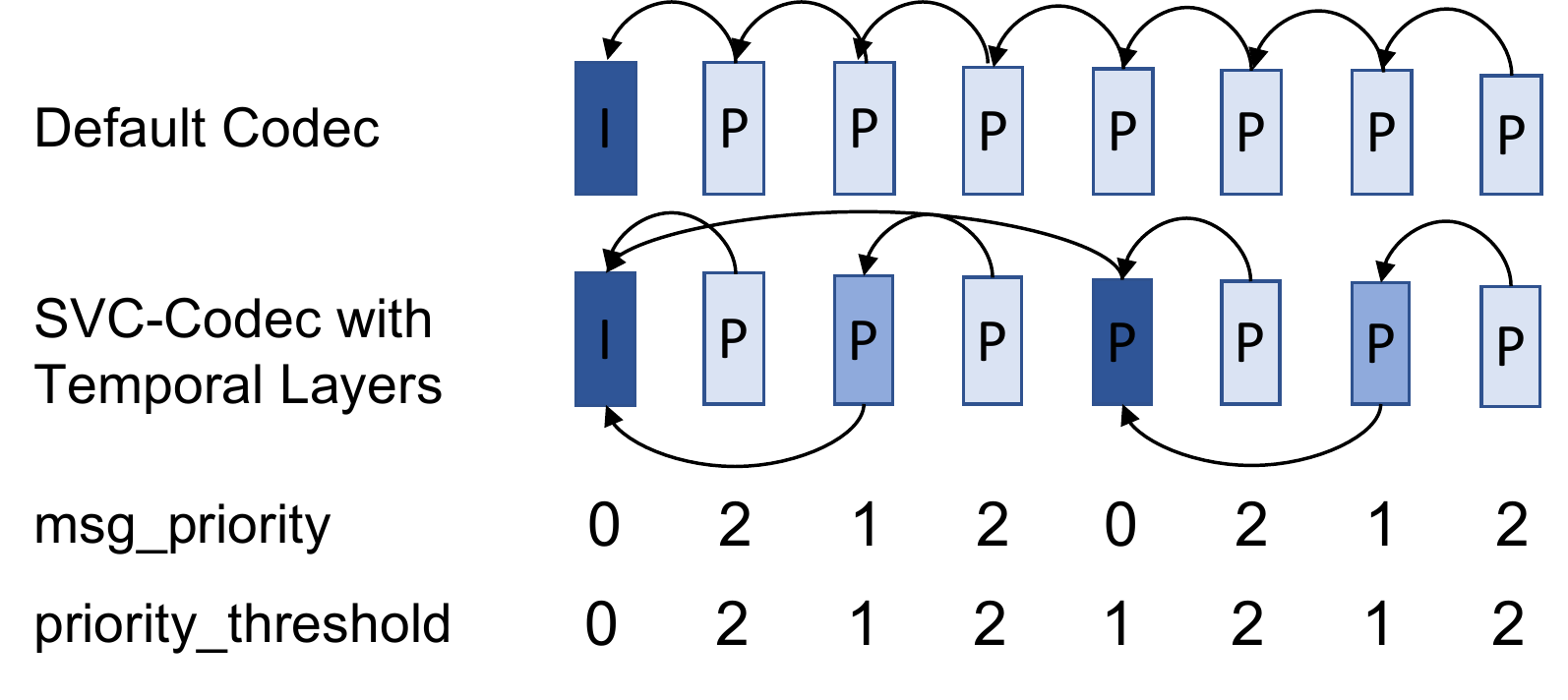}
\vspace{-10pt}
\caption{Single-layered video codec (top) and SVC with three temporal layers (bottom).}
\vskip -0.2 in
\label{fig:freshcodec}
\end{figure}

\subsection{Real-time Video with Frame Rate Adaptation}
\label{sec:casestudy1} 


Our first case study considers the requirement where the freshest video frame must be delivered at fixed quality (e.g. for applications that require processing the received video using an ML algorithm). 

\paragraphb{Video Encoding.} Commonly used video codecs (e.g. H.264~\cite{h264} and VP9~\cite{vp9}) encode video frames in chunks called ``group of pictures'' (GoP).  As shown in Figure~\ref{fig:freshcodec} (top), the first frame in each GoP (the I-frame) is intra-coded and can be decoded independently. Each subsequent P-frame only encodes the delta from the previous frame. The successful decoding of a P-frame at the receiver therefore requires successful delivery of all previous frames in the GoP. This limits the tolerance to in-network packet drops.

To better exploit \oursys, we make use of multiple temporal layers supported by scalable video codec(SVC)\cite{h264-svc, vp9-spatial, webrtc-svc}. Here, we apply VP8-SVC with three temporal layers. Figure~\ref{fig:freshcodec} (bottom) shows the dependency structure between frames. Frames marked with priority value $i$ serve as reference only for frames with priority value $j > i$, and can be dropped upon the arrival of a new frame with priority\_threshold $k \le i$, without affecting any subsequent frames. The usage of VP8-SVC introduces slight overhead -- the average frame size is 15\%-18\% larger than that encoded in default VP8. Our results show how \oursys improves overall performance in spite of these overheads, in comparison with baselines that use the default codec.




\paragraphb{Dropping Policy.} Using the video codec described above, we treat each frame as a single message, and only make use the \primone primitive. We set the drop\_flag in every message. Figure~\ref{fig:freshcodec} shows the msg\_priority, and priority\_threshold that we set for each message. The bitrate\_threshold is set to the default value of zero to disable \primtwo primitive.

\paragraphb{Baselines.} We compare \oursys with two state-of-the-art schemes for real-time video transmissions: (i) AWStream~\cite{awstream}, configured to adapt frame rate based on the bandwidth estimated by the underlying transport (we experiment with both TCP Cubic and BBR). We use the default VP8 encoding for the video. (ii) Salsify~\cite{salsify}, modified to solely tune the frame rate keeping the frame quality fixed (we refer to this as Salsify$^*$). As in the original Salsify design, it uses Sprout's EWMA-based congestion control mechanism (a more aggressive variant of Sprout~\cite{sprout}), and a functional codec based on VP8 encoding. We encode the videos for all three schemes with a fixed quantization level of 17. 

\paragraphb{Metric.} We use ``age of information'' (AoI)~\cite{aoi} to capture freshness as the time elapsed since the latest frame delivered at the receiver application was sent out by the sender application (lower AoI implies higher freshness). 
To capture the worst-case freshness, we measure the AoI just before a frame is received, and compute the 99\%ile and median values across all frames in a video. AoI can be high either due to high queuing delays or low frame rate (or both).

\paragraphb{Results.}
As shown in Figure~\ref{fig:eval-usecase1}, \oursys outperforms AWStream and Salsify$^*$, in spite of using a less efficient encoding strategy. 
The tail AoI with \oursys is 1.6-18$\times$ lower than that of AWStream.
We found that AWStream (designed for more stable WAN bandwidth) is overly conservative in increasing its sending rate (upgrading to a higher throughput configuration only after its application buffer has been empty for 2s).
The tail AoI with \oursys is 1.5-2$\times$ lower than Salsify$^*$'s - Salsify$^*$ suffers from higher tail latencies and lower link utilization due to slower reaction to changing network capacity as compared to \oursys.

\subsection{Real-time Video with Quality Adaptation}
\label{sec:casestudy2} 
Many real-time apps (e.g. video conferencing and live streaming) allow tuning the quality of the video content to sustain stable FPS with low latency. Our second case study considers this requirement.

\begin{figure}[t!]
\centering
\includegraphics[width=0.45\textwidth]{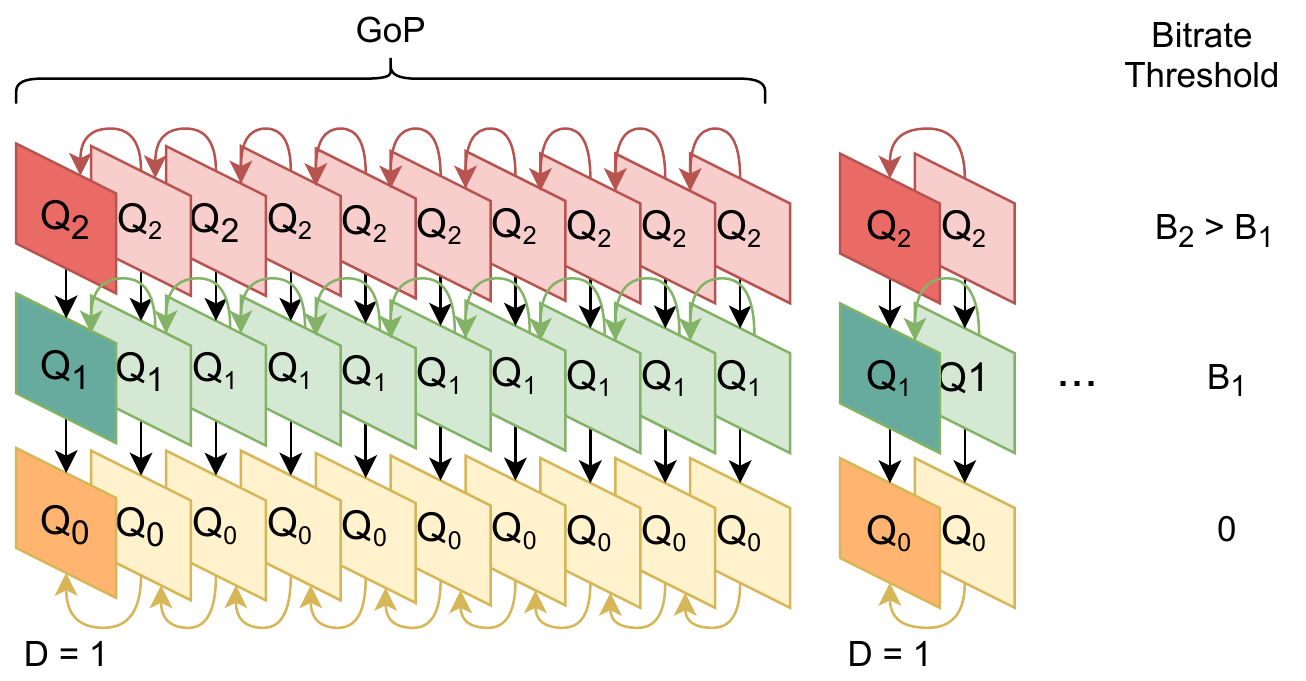}
\vskip -0.15 in
\caption{SVC encoding with three quality layers, and the dropping primitive parameters.} 
\vskip -0.15 in
\label{fig:usecase2_codec}
\end{figure}

\paragraphb{Video Encoding.} We exploit support for multiple spatial (quality) layers in SVC for quality adaptation with \oursys.~\footnote{Future work can explore using other layered codecs, e.g. neural video codec\cite{neural-codec}.} Figure~\ref{fig:usecase2_codec} shows how frames are encoded in SVC with 3 spatial layers (depicted as $Q_0$, $Q_1$, and $Q_2$ from lowest to highest). The decoding of a higher layer frame depends on the successful decoding of all lower layers and the corresponding layer of the previous frame.

In our SVC-based \oursys application, we fix the number of quality levels to 3, and use the bandwidth estimation ($B$) provided by the underlying \oursys transport to dynamically adjust the encoding quality levels for each GoP.~\footnote{We found that a frame with up to three SVC layers can be decoded within 10ms using a single-threaded code.} More specifically, quality $Q_2$ is configured such that its cumulative target bitrate is $B_2\approx B$, $Q_1$ is configured to a cumulative bitrate of $B_1\approx 0.5B$, and $Q_0$ is configured to a bitrate $B_0\approx 0.2B$.
We reduce the GoP size to 10 frames. This introduces a slightly higher overhead, but allows faster switching across quality levels.


\paragraphb{Dropping Policy.} We use the \primtwo primitive, treating each layer of each frame as an individual message.
We mark the messages corresponding to quality $Q_0$, $Q_1$ and $Q_2$ with bitrate\_thresholds of zero, $B_1$ and $B_2$ respectively.

Overall, this policy achieves the desired trade-off in quality, throughput and message latency. However, it has a few caveats: (1) It may result in high queuing when the bandwidth is lower than $B_0$ (the required bitrate for the lowest quality frames). We handle this by using our \primone primitive, and setting the drop\_flag in the first $Q_0$ message in each GoP. With priority\_threshold of zero, this triggers a drop in all queued up frames of the previous GoP. 
(2) There might be scenarios where the bandwidth increases after some of the $Q_2$ or $Q_1$ messages towards the beginning of a GoP have been dropped. Transmission of $Q_2$ messages in the GoP that are then enqueued is wasteful as they cannot be decoded at the receiver. Nonetheless, we observe that if the bandwidth is high enough to sustain their bitrate, transmission of these messages do not generally block transmission of newer frames in the stream. Moreover, by limiting the GoP to 10 frames, we limit the scope of such wasted frame transmissions.

\begin{figure*}[t!]
    \centering
    \subfigure[\normalsize{ATT}]{\includegraphics[width=0.32\textwidth]{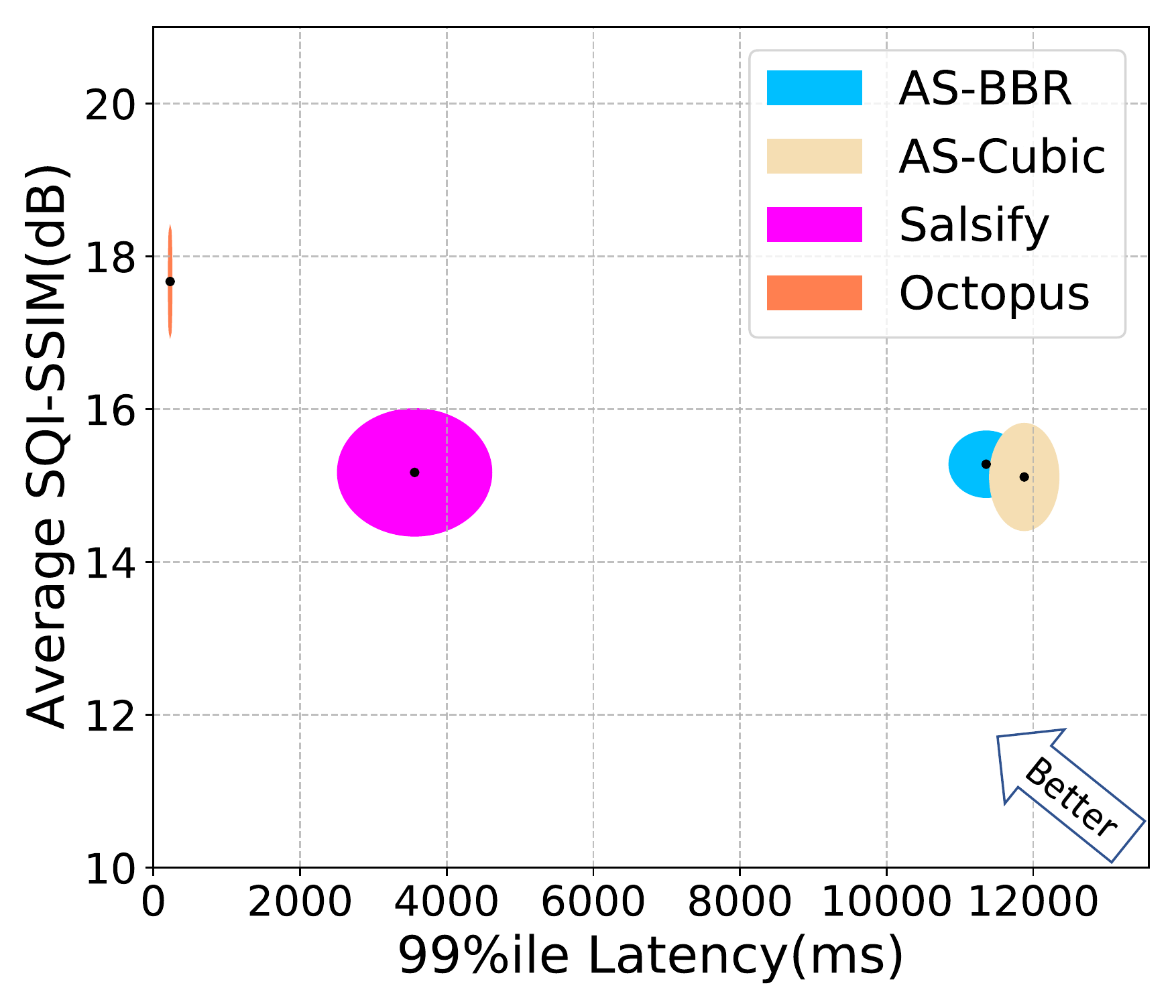}}\quad
    \subfigure[\normalsize{TMobile}]{\includegraphics[width=0.32\textwidth]{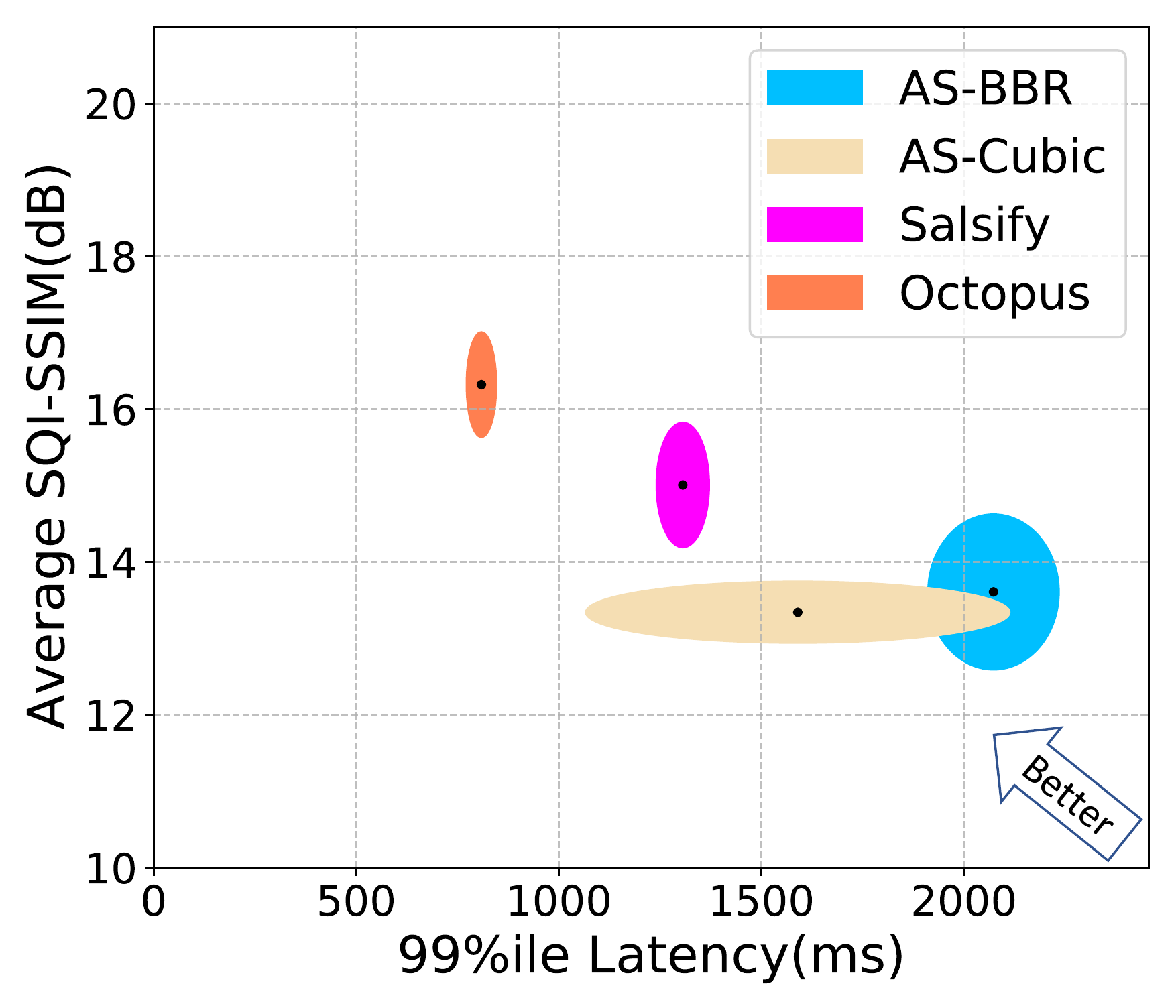}}\quad
    \subfigure[\normalsize{Verizon}]{\includegraphics[width=0.32\textwidth]{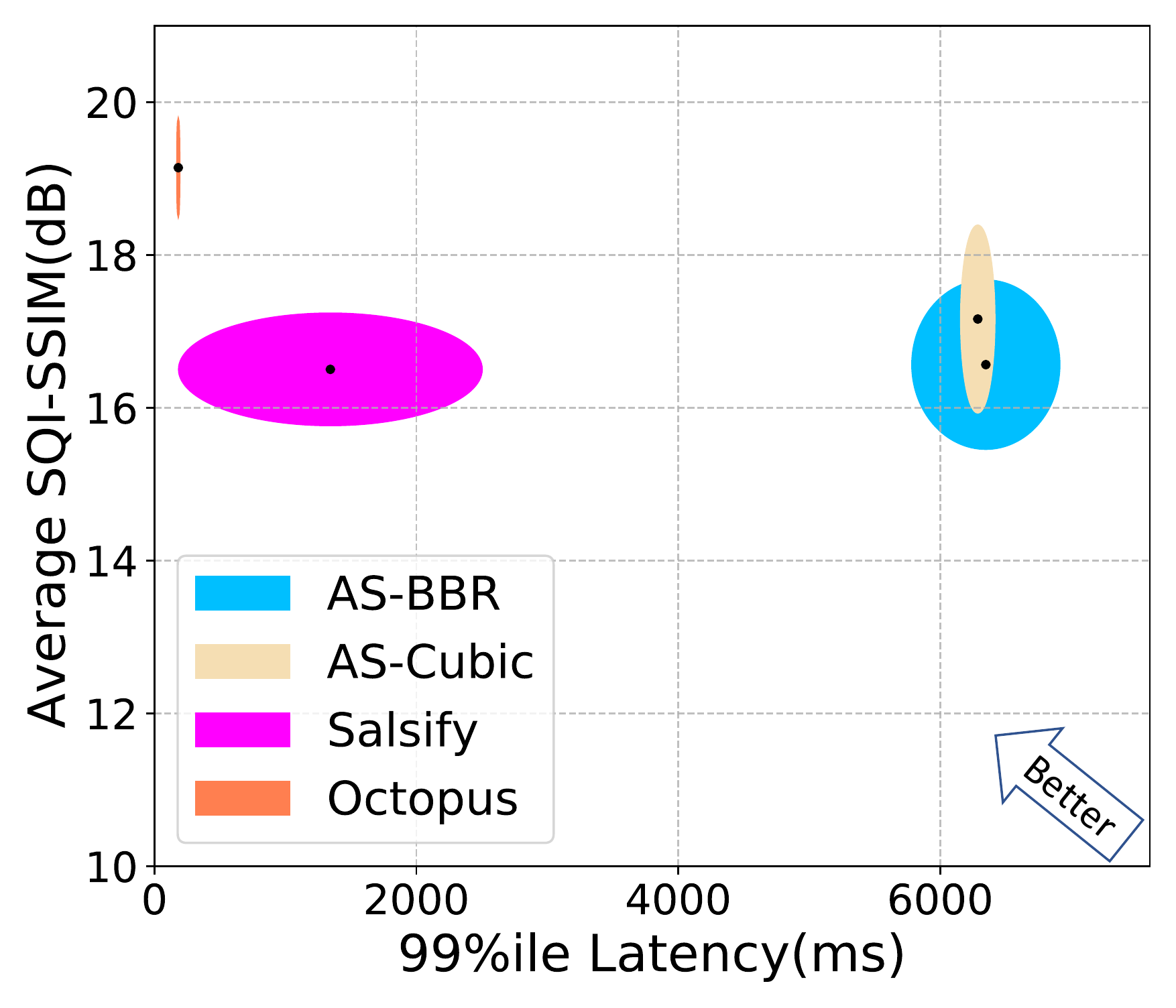}}
    \vskip -0.1 in
    \caption{The 99\%ile tail latency v.s. video quality of Octopus and other baselines in different LTE download traces with RTT 60ms. The axes of the ellipse reflect the standard deviations in SQI-SSIM and tail latency.}
    \label{fig:eval-usecase2}
\end{figure*}

\paragraphb{Baselines.} We compare Octopus for this usecase with: (i) AWStream, configured this time to adapt video quality, (ii) Unmodified Salsify.

\paragraphb{Metric.} We measure the video quality using SQI-SSIM\cite{sqi-ssim, vantage} -- it computes the quality of each decoded frame using structural similarity(SSIM)~\cite{ssim, ssim-psnr}, and the quality of each undecoded (or dropped) frame as the product of the SSIM of the last available frame and an exponential decay function. The resulting video QoE score is the average quality across each (decoded and undecoded) frame that was sent by the sender app. We also record the latency of each delievered frame. It's desirable to achieve high SQI-SSIM and low latency.

\paragraphb{Results.}
As shown in Figure~\ref{fig:eval-usecase2}, \oursys achieves higher average SQI-SSIM with lower tail latency than the baselines (the median latency, shown in Appendix, is similar across different schemes). \oursys achieves 14-31\% higher SQI-SSIM than AWStream, with 2-50$\times$ lower tail latency. The difference in SQI-SSIM stems from AWStream's conservative behaviour when upgrading to higher throughput configuration. AWStream's tail latency is relatively high due to poor reaction when link capacity suddenly drops to very low values. \oursys achieves 17\% higher SQI-SSIM than Salsify, with 1.6-10$\times$\ lower tail latency. The difference in SQI-SSIM largely stems from the inefficiency of the functional VP8 codec used in Salsify. The difference in tail latency stems from \oursys's faster reaction to sudden drops in bandwidth. 
 

\begin{figure}[t]
\centering
\includegraphics[width=0.40\textwidth]{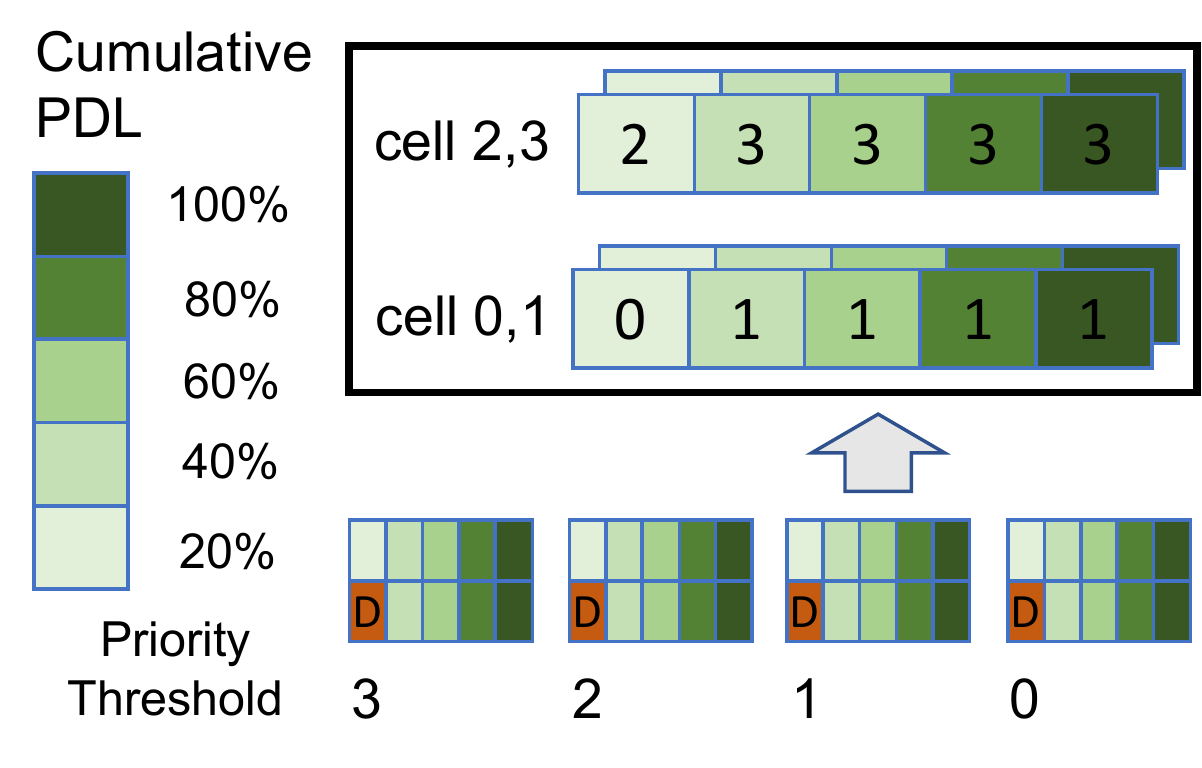}
\vskip -0.1 in
\caption{The upper graph shows a volumetric video frame with four cells, and each cell with five layers. The lower graph shows dropper messages and their priority thresholds.}
\vskip -0.2 in
\label{fig:usecase3_codec}
\end{figure}

\subsection{Real-time Volumetric Video Streaming}
\label{sec:casestudy3} 
Real-time volumetric streaming enables viewers to watch videos in six degree of freedom (6DoF), and is becoming popular in education, entertainment, and healthcare. We look at how \oursys can adapt the quality of bandwidth-intensive volumetric streams to minimize delay while trying to sustain a stable frame rate. 

\paragraphb{Video Encoding.} Uncompressed volumetric video frames are represented as point clouds which record the attributes such as coordinates and colors of every point. A frame is often spatially segmented into multiple cells, and each cell can be independently encoded and decoded~\cite{vivo, groot}. This segmentation enables the server to stream a subset of cells or adapt the point density level (PDL) of cells based on the viewer's current viewport to significantly reduce network bandwidth requirement.

Figure~\ref{fig:usecase3_codec} shows how we encode a volumetric frame. In our experiments, every video frame is segmented in four cells. Instead of encoding every cell with a specific PDL, we further divide the origin cell into five layers, each comprising 20\% points and encoding every layer independently, so that Octopus can safely drop specific layers in a cell to adapt its PDL. 
This results in only 10\%-12\% overhead in terms of bandwidth usage which, as our results show, is more than compensated by \oursys' fast reaction to bandwidth changes compared to the baseline. 

\paragraphb{Dropping Policy.}
Based on the Occlusion Visibility(OV) and Distance Visibility(DV) adaptive streaming approaches from ViVo~\cite{vivo}, we first drop layers of occluded cells, and then drop higher density layers of non-occluded cells, as queues start building up. For simplicity we assume the viewport is fixed and cell 0 and cell 1 are front cells and cell 2 and cell 3 are occluded cells (we can dynamically update the cells priority for changing viewport using the viewport movement and prediction model from ViVo). 

We apply \primone primitive, and treat each layer inside a cell as an individual message. The message priority for each such layer is indicated in Figure \ref{fig:usecase3_codec}. We set the drop\_flag in the first layer of the non-occluded cell 0 in every frame. The drop\_thresholds in these dropper messages are set to a repeating pattern of 3, 2, 1, and 0, which enables the smooth adaptation of cell PDLs in buffer.

\paragraphb{Baselines.}
We compare Octopus with ViVo in this usecase. Since the source code of ViVo is unavailable, we implement a version to our best knowledge. Here, the ViVo client uses a throughput-based rate adaptation algorithm\cite{festive} to estimate the link capacity, and determine the PDL of occluded cells and front cells to fetch.

\paragraphb{Metric.}
Similar to \S\ref{sec:casestudy2}, we use SQI-SSIM to compute the quality of received volumetric video frames, and 99\% frame delivery tail latency to capture freshness of frames. To calculate the SSIM, we leverage the mappings from cell PDLs to SSIM perceived by viewers at a given distance developed by ViVo. In the experiment, the default distance between the viewport and cells is 2.5m.

\begin{figure}[t]
    \centering
    \subfigure[\footnotesize{5G trace1}]{\includegraphics[width=0.23\textwidth]{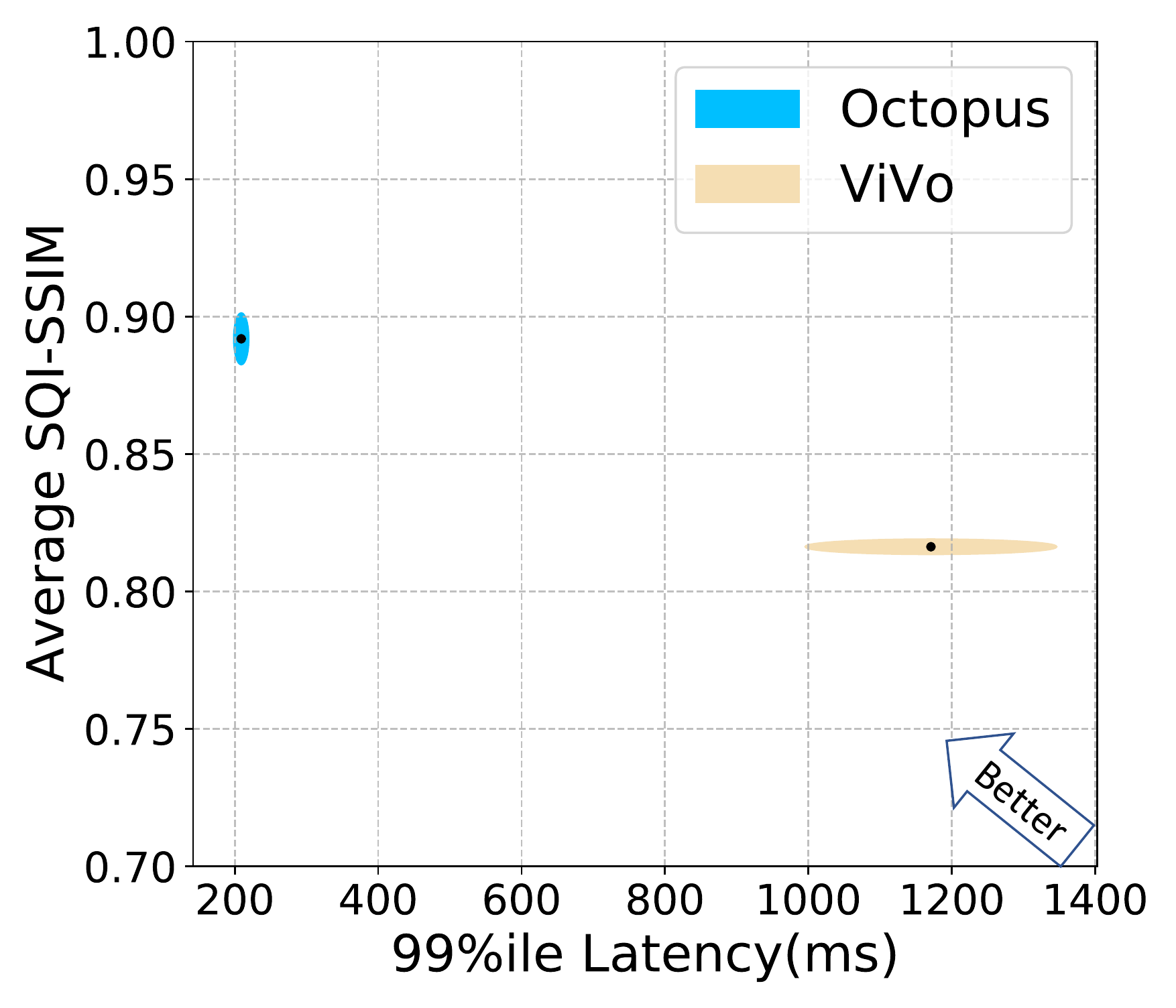}}
    \subfigure[\footnotesize{5G trace2}]{\includegraphics[width=0.23\textwidth]{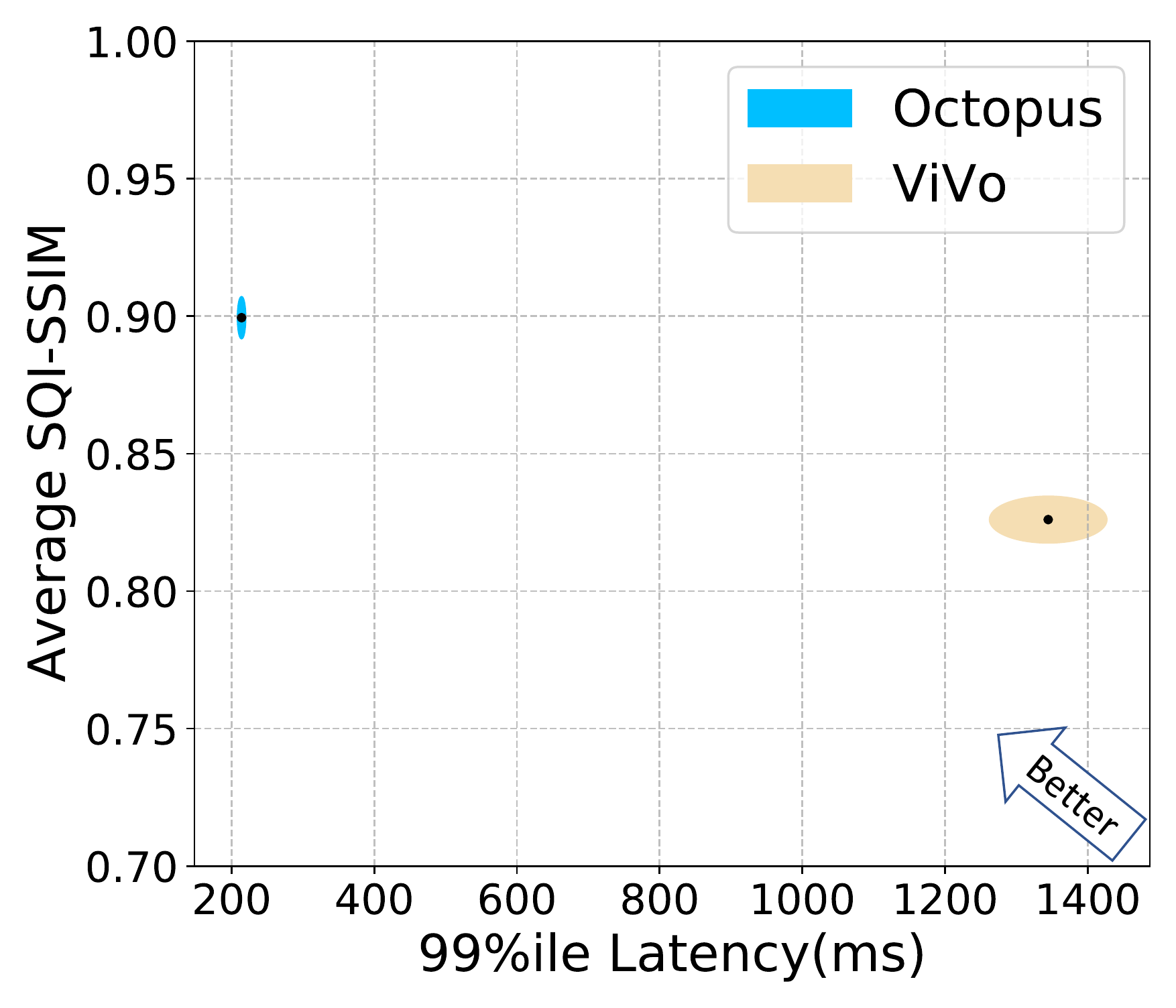}}
    \caption{The 99\% tail latency v.s. volumetric video quality of Octopus and ViVo in two 5G traces with RTT 60ms.}
    \vspace{-10pt}
    \label{fig:eval-usecase3}
\end{figure}

\paragraphb{Results.}
From Figure~\ref{fig:eval-usecase3}, Octopus achieves 9\% percent higher average SQI-SSIM  and 83\% lower tail latency than ViVo. The reason that ViVo has lower frame quality stems from its conservative rate adaptation algorithm, which uses the harmonic mean of the download rate of previous 20 video frames as the capacity estimate for the next frame. ViVo's tail latency is relatively high because TCP reliably delivers all packets even when the real network bandwidth suddenly drops below the estimate. 

\label{sec:volumetric} 
\section{Detailed Evaluation}
\label{sec:eval-details}

\begin{figure*}[t!]
    \centering
    \subfigure[\footnotesize{Comparison with OctoBBR (Octopus without in-network support).}]{\includegraphics[width=0.23\textwidth]{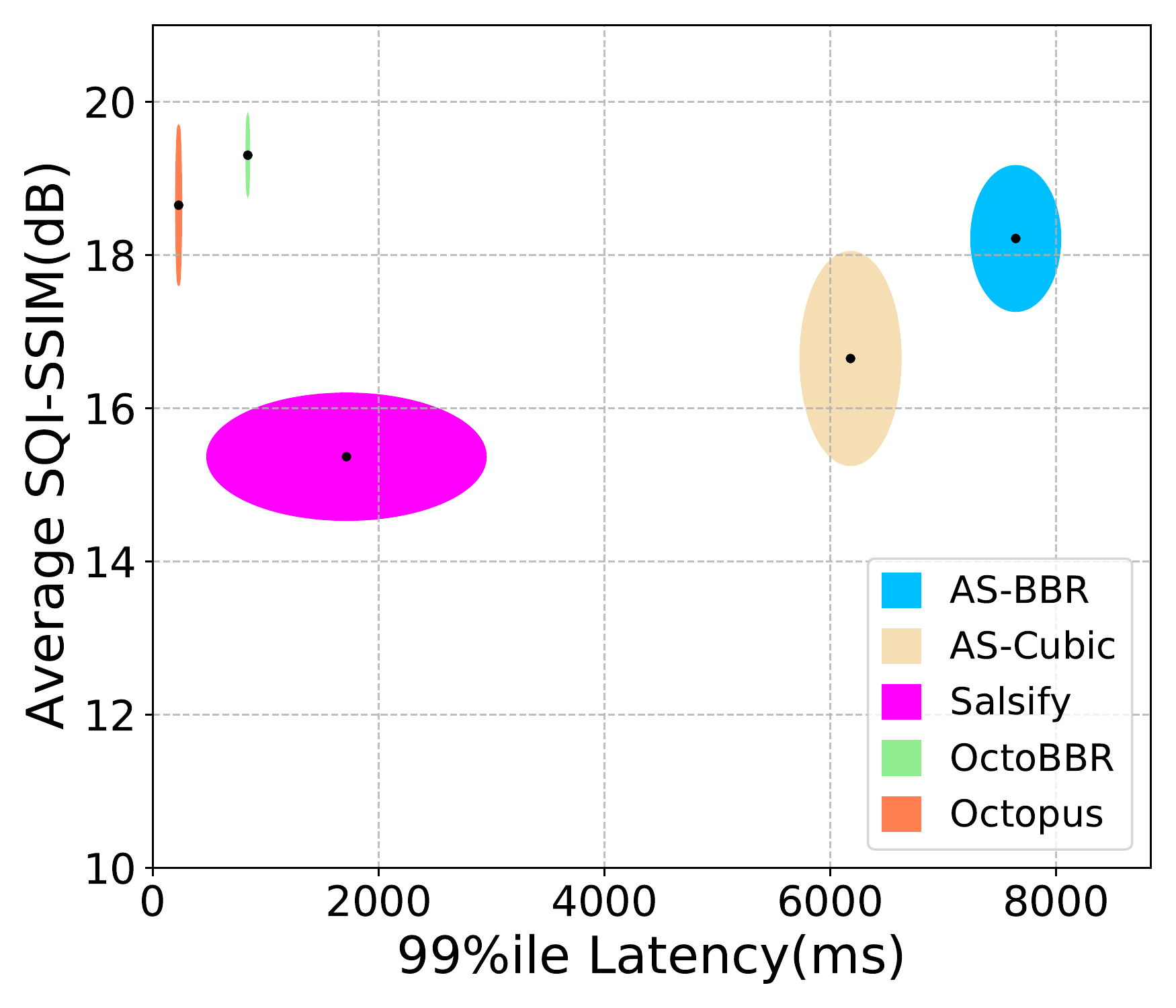}\label{fig:octobbr}}
    \quad
    \subfigure[\footnotesize{Comparison with priority-based dropping at different buffer sizes.}]{\includegraphics[width=0.23\textwidth]{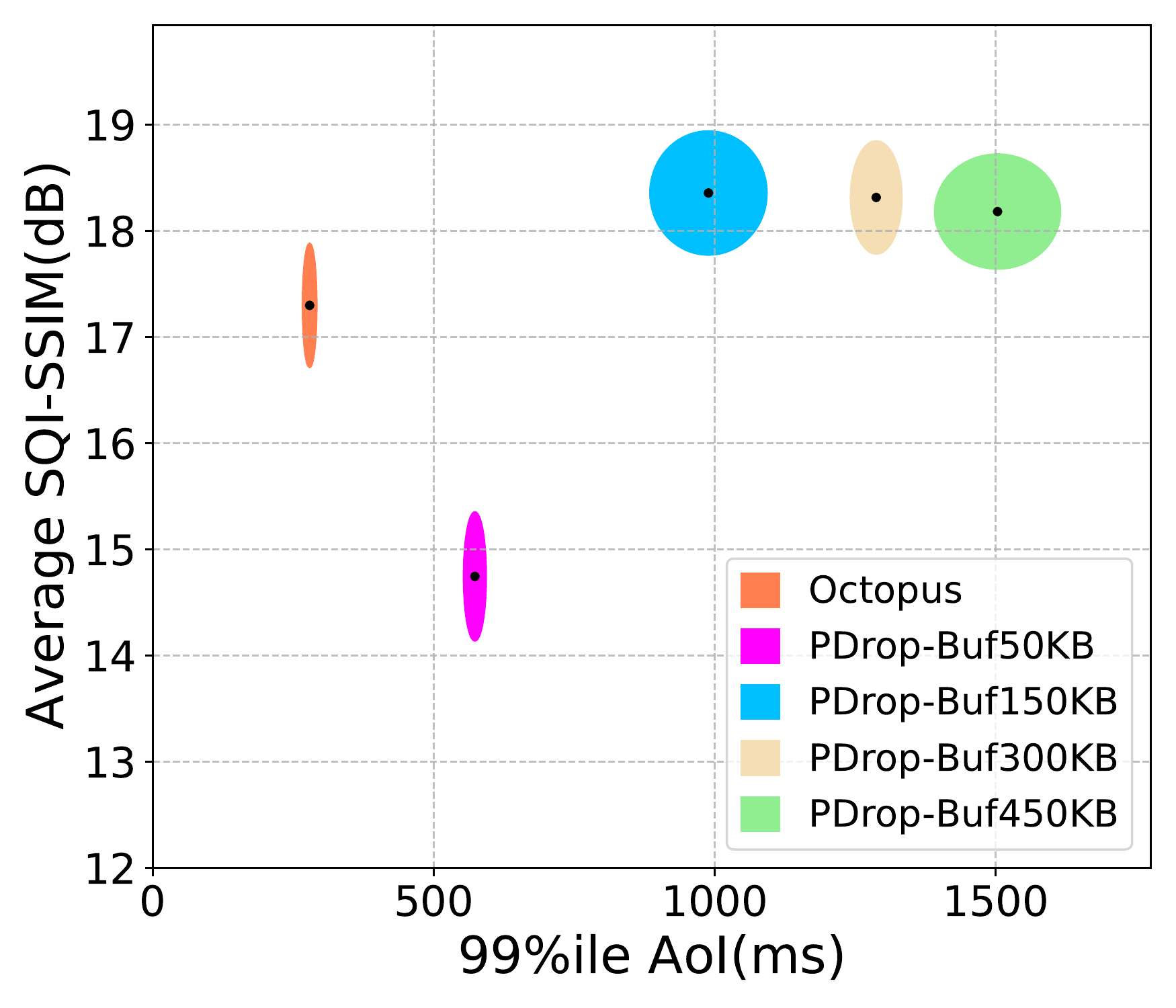}\label{fig:octopus-anet}}
    \quad
    \subfigure[\footnotesize{When competing with a backlogged flow on a legacy switch.}]{\includegraphics[width=0.23\textwidth]{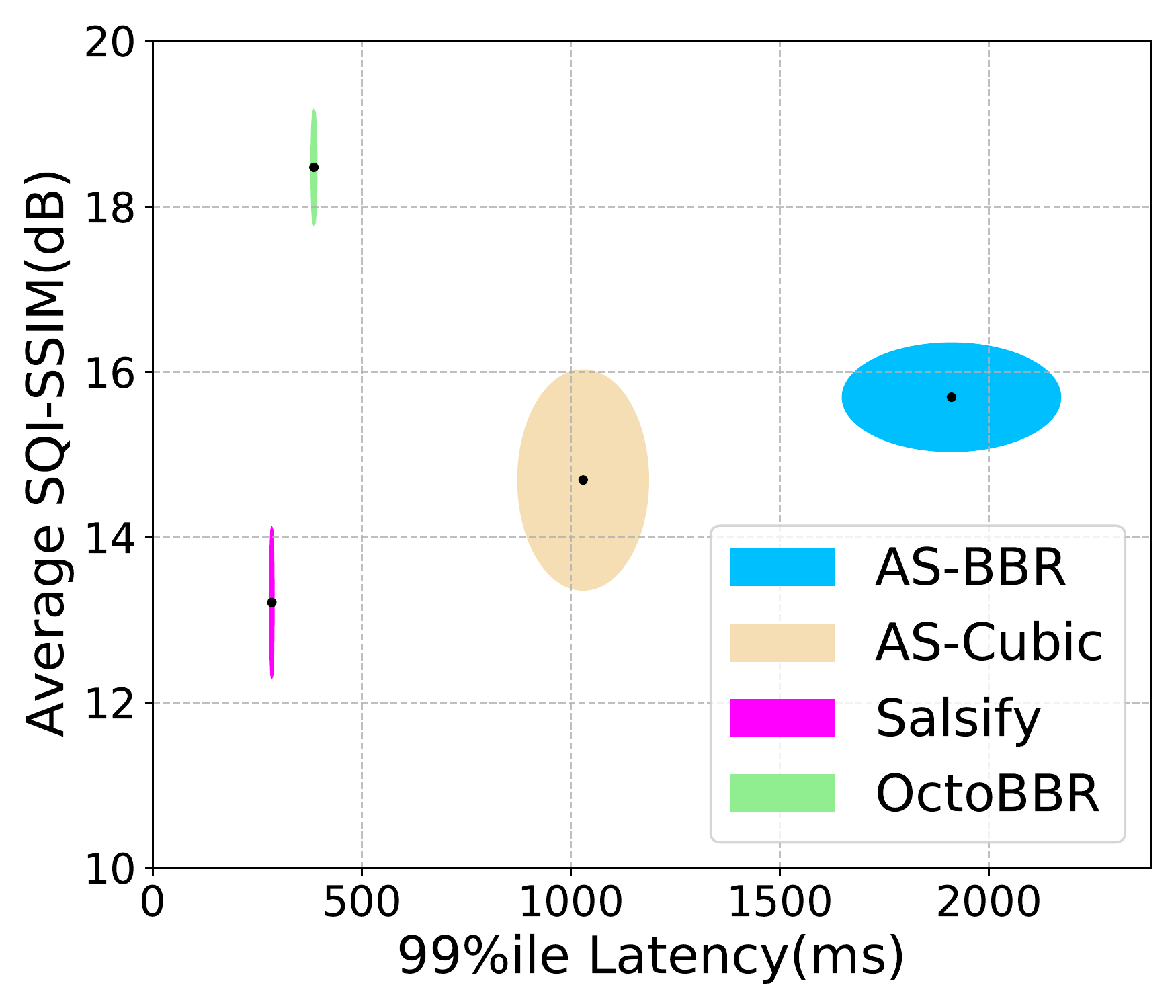}\label{fig:bottleneck-compete}}
    \quad
    \subfigure[\footnotesize{Two competing video streams (indicated by different markers) }]{\includegraphics[width=0.23\textwidth]{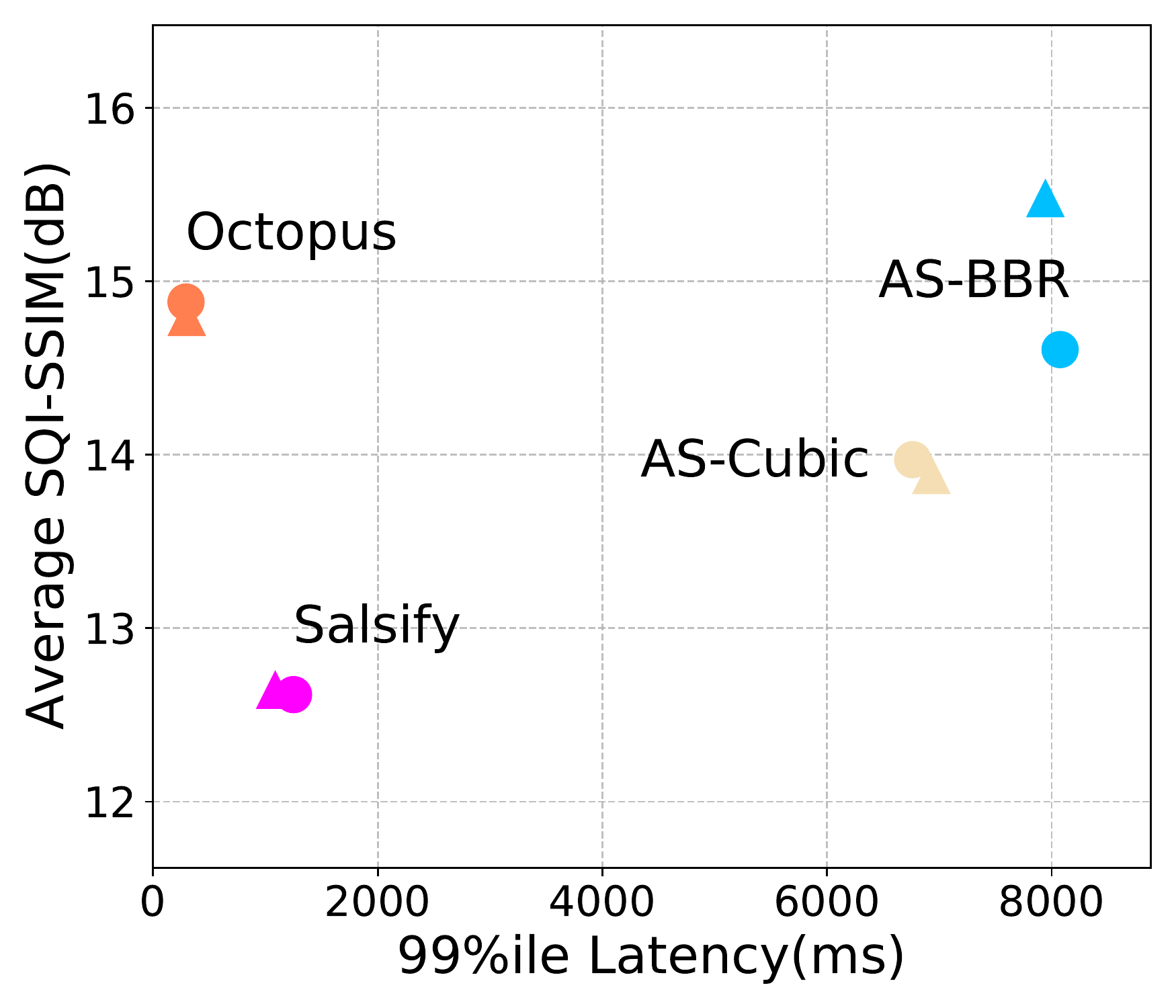}\label{fig:bottleneck-multi}}
    \vskip -0.1 in
    \caption{The 99\%ile tail latency v.s. video quality of Octopus and other baselines for real-time video under different scenarios on the Verizon trace with 120ms RTT.}
\end{figure*}

We use our case-study in 
\S\ref{sec:casestudy2} for a more in-depth evaluation of \oursys. For brevity, we only present results with Verizon download trace at RTT 120ms (we see similar trends with other traces). We use our srsRAN testbed for the results in \S\ref{sec:octobbr}, and Mahimahi emulator (that allows for greater configurability in experimental scenarios and baselines) for the remaining experiments.

\subsection{Decoupling the impact of \oursys Endpoint Logic}
\label{sec:octobbr}

We now evaluate the performance impact of \oursys endpoint's logic in isolation. For this, we use \oursys transport protocol as it is, but disable the Octopus router logic. This effectively models BBR using Octopus logic for transport buffer management and app content adaptation. We refer to this scheme as \OctoBBR.

Figure~\ref{fig:octobbr} compares \OctoBBR with \oursys, as well as the baselines described in \S\ref{sec:casestudies}. There are two key takeaways from these results:

\paragraphi{(i)} \oursys performs better than \OctoBBR. 
\oursys achieves a more desirable trade-off than \OctoBBR 
with 75\% lower tail latency and only 3\% lower SQI-SSIM. This shows that in-network content adaptation (based on more timely and accurate knowledge of link capacity and queue build-up) is useful.  

\paragraphi{(ii)} \OctoBBR performs better than the state-of-the-art endpoint-based solutions (Salsify and AWStream). This highlights the benefits of direct content adaptation at the endpoint's transport buffer by dropping messages using \oursys' primitives when there's a mismatch between the application sending rate and the pacing rate of transport protocol.


\subsection{Comparison with Priority-based Message Dropping}

We next compare \oursys' in-network message dropping logic with a simpler priority-based message dropping mechanism (inspired from~\cite{bhattacharjee1998network,zegura-an}). With this mechanism, when the router buffer is full, and a new message with priority $p$ arrives, instead of dropping the incoming message, the router drops all queued up messages with lower priority (i.e. having priority value $> p$). We set highest priority $0$ for base ($Q_0$) layer, followed by priority $1$ for $Q_1$, and then priority $2$ for $Q_2$.

Figure~\ref{fig:octopus-anet} compares \oursys with the priority-based dropping scheme described above (represented as PDrop). We fix the buffer size to 375KB for \oursys, and use varying buffer sizes for PDrop. 
We find that PDrop is sensitive to buffer sizes. Larger buffer size results in fewer message drops, and larger latency. A small buffer size, on the other hand, results in significantly lower video quality. Packet delay thresholds or deadlines would be similarly difficult to tune. 
\oursys primitives are able to react faster and more appropriately by triggering drops directly based on message arrival and link capacity, instead of relying on a buffer threshold. This allows \oursys to achieve a more desirable trade-off.  


\subsection{Competing Backlogged Flow at a Legacy Switch}
\label{sec:bottleneck}



We emulate a legacy (non-\oursys) switch with a static link bandwidth of 12Mbps. In addition to the real-time stream, we generate a competing backlogged TCP (BBR) flow that shares the switch buffer. Figure~\ref{fig:bottleneck-compete} presents the 
results of \oursys and comparative baselines in this scenario. With the \oursys router logic disabled for the legacy switch, only the \oursys endpoint logic kicks in (which, in line with \S \ref{sec:octobbr} is represented as \OctoBBR). 
Overall, we find that \OctoBBR achieves the most desirable trade-off between SQI-SSIM and latency. We find that Salsify (using Sprout as the underlying transport) competes poorly with the backlogged TCP flow, utilizing less than its fair share of the link capacity, which in turn reduces the video quality. \OctoBBR competes fairly with the backlogged flow, allowing it to achieve 40\% higher SQI-SSIM than Salsify. The graph comparing the link utilizations achieved by Salsify and \OctoBBR is shown in Appendix~\ref{appendix-link}. Both Salsify and \OctoBBR see similar increase in in-network delay due to the competing flow, but \OctoBBR sees slightly higher self-inflicted delay as it is able to send more messages into the network (that the legacy switch does not drop). This results in 18\% higher tail latency in \OctoBBR compared with Salsify. AWStream has lower video quality (due to the aforementioned conservative app behavior, which results in low link utilization) and larger tail latency (due to higher queuing delay in the endpoint's transport buffer). 

We also evaluated a scenario with multiple bottlenecks, one at a legacy switch and another at an \oursys-enabled cellular base-station. We present those results in Appendix~\ref{appendix-multiple}.

\subsection{Competing real-time streams sharing \oursys buffer}
\label{sec:multi-streams}

We next evaluate a scenario with two real-time video streams sharing the same router queue. For the \primtwo condition, the router computes the max-min fair rate $BW_i$ for each stream $i$ from the observed per-stream enqueuing rate and the overall rate at which the queue is served. We use the policy from \S\ref{sec:casestudy2} for both streams. Figure~\ref{fig:bottleneck-multi} compares \oursys with Salsify and AWStream (the result for each video stream is denoted using two different markers). We find that \oursys achieves higher SQI-SSIM and lower tail latency for both video streams, as compared to Salsify and AWStream. This shows that \oursys can appropriately handle multiple real-time streams, when the dropping parameters in each stream are configured to  minimize the self-inflicted delay.


\section{Scope and Limitations}
\label{sec:limitations} 

The goal of our work was to present a new design point for controlling congestion that is based on in-network content adaption, and show the promise of this approach. We acknowledge that our approach requires changes at both the endpoints and the cellular routers, making it difficult to deploy immediately. We list a few other limitations of our work below:

\paragraphb{Applicability.} Our approach does not entirely replace the need for feedback-based controllers -- it is well-suited only for those apps that can support in-network content adaption. 

\paragraphb{On generality of our primitives.} 
While our primitives can seemingly capture a range of requirements for real-time apps, we are yet to formally analyze their expressiveness.

\paragraphb{Coordination via packet headers.} \oursys needs 12 bytes of header space to convey the per-message dropping parameters. While our prototype uses application-layer headers, an actual deployment may need to use other alternatives (e.g. IP options/extensions). Diving into the feasibility of these alternatives is beyond the scope of our work. 

\paragraphb{Competition with buffer-filling cross-traffic.} 
We assume that a user's real-time traffic is isolated from their non-realtime traffic using standard mechanisms~\cite{downlink-schedule, lte-qos}. We find that in the absence of such isolation, \oursys' dropping logic competes poorly with flows that aggressively fill up the network buffer.~\footnote{Disabling \oursys's in-network logic and relying on its transport logic still outperforms state-of-the-art baselines in such scenarios, as discussed in \S\ref{sec:bottleneck}.} Such a fate is fundamental to any scheme that is designed to react fast to bandwidth variations, and isolation from buffer-filling cross-traffic is therefore a common assumption that is also made by prior work~\cite{sprout, salsify, abc}.

\section{Conclusion and Discussion}
\label{sec:discussion}

This paper presented \oursys, a system designed to achieve high throughput and low latency for real-time transmissions over cellular networks. Typical real-time apps adapt their content (data quality and frame rate) based on the network bandwidth estimated by the endpoint transport. \oursys allows these apps to send data aggressively, and to instead specify how content can be adapted by the lower layers using per-message parameters. \oursys transport and router logic implement generalized primitives to drop messages as per app-specified parameters when a queue build ups or link bandwidth reduces. This allows \oursys to react in a timely manner, and perform significantly better than state-of-the-art. 

Our work leaves several interesting directions open for future research. For example, how can application frameworks be designed to better exploit a system like \oursys? Can this approach be extended to other scenarios, e.g. for datacenter workloads? How well do \oursys' primitives generalize to other usecases? etc.

\bibliographystyle{ACM-Reference-Format}
\bibliography{paper}

\section*{Appendix}
\begin{appendices}
\clearpage

\section{Dropping primitives in P4}
\label{appendix-p4}
As a proof-of-concept to show implementation feasibility of Octopus's dropping primitives in hardware, we implement \oursys' dropping primitives in P4\cite{p4} bmv2.
\oursys maintains a table of latest dropper messages, indexed by the priority threshold. It refers to this table at the egress to determine whether the packet should be dropped or forwarded. This table must be updated when a dropper message arrives at the ingress. However, P4 does not allow sharing memory between ingress and egress. We resolve this issue by cloning the tail packet of the dropper message at the ingress and sending this clone in high priority. This allows the egress to dequeue the clone and update the drop table before dequeuing regular data packets sent at a lower priority. The egress drops the clone after updating the table. Algorithm\ref{algo2} is the pseudocode for the P4 implementation logic.

\begin{algorithm}[]
\caption{Semantic packet drops in P4}
\begin{algorithmic}
\Procedure{Ingress}{packet}
\If {packet.hasDropFlag() \textbf{and} packet.isTail()}
    \State packet.priority() $\gets$ HIGH\_PRIO
    \State Clone(packet)
\EndIf
\State packet.priority() $\gets$ LOW\_PRIO
\State \Return Enqueue(packet)
\EndProcedure

\Procedure{Egress}{packet}
\State \textbf{Register} reg\_droppers[NUM\_PRIORITY]
\State \textbf{Register} reg\_current\_drop[1]
\If {packet.isClone()}
    \State threshold $\gets$ packet.priorityThreshold()
    \State reg\_dropers[threshold] $\gets$ packet.msgID()
    \State \Return Drop(packet)
\EndIf
\State msgid $\gets$ packet.msgID()
\State prio $\gets$ packet.priority()
\If {packet.isHead()}
    \State drop\_msgno $\gets$ max(reg\_droppers[0]...reg\_droppers[prio])
    \State isdrop $\gets$ ( msgid $<$ drop\_msgno
    \State \hspace*{2em} \textbf{or} packet.bitrate() $>$ dequeueRate() )
    \If {isdrop}
        \State reg\_current\_drop[0] $\gets$ msgid
    \EndIf
\EndIf

\If {reg\_current\_drop[0] $=$ msgid}
    \State \Return Drop(packet)
\EndIf
\State \Return Dequeue(packet)
\EndProcedure
\end{algorithmic}
\label{algo2}
\end{algorithm}

\section{Supplementary case study results}
\label{appendix-rtt120}
\begin{figure*}[h!]
    \centering
    \subfigure[\normalsize{ATT}]{\includegraphics[width=0.32\textwidth]{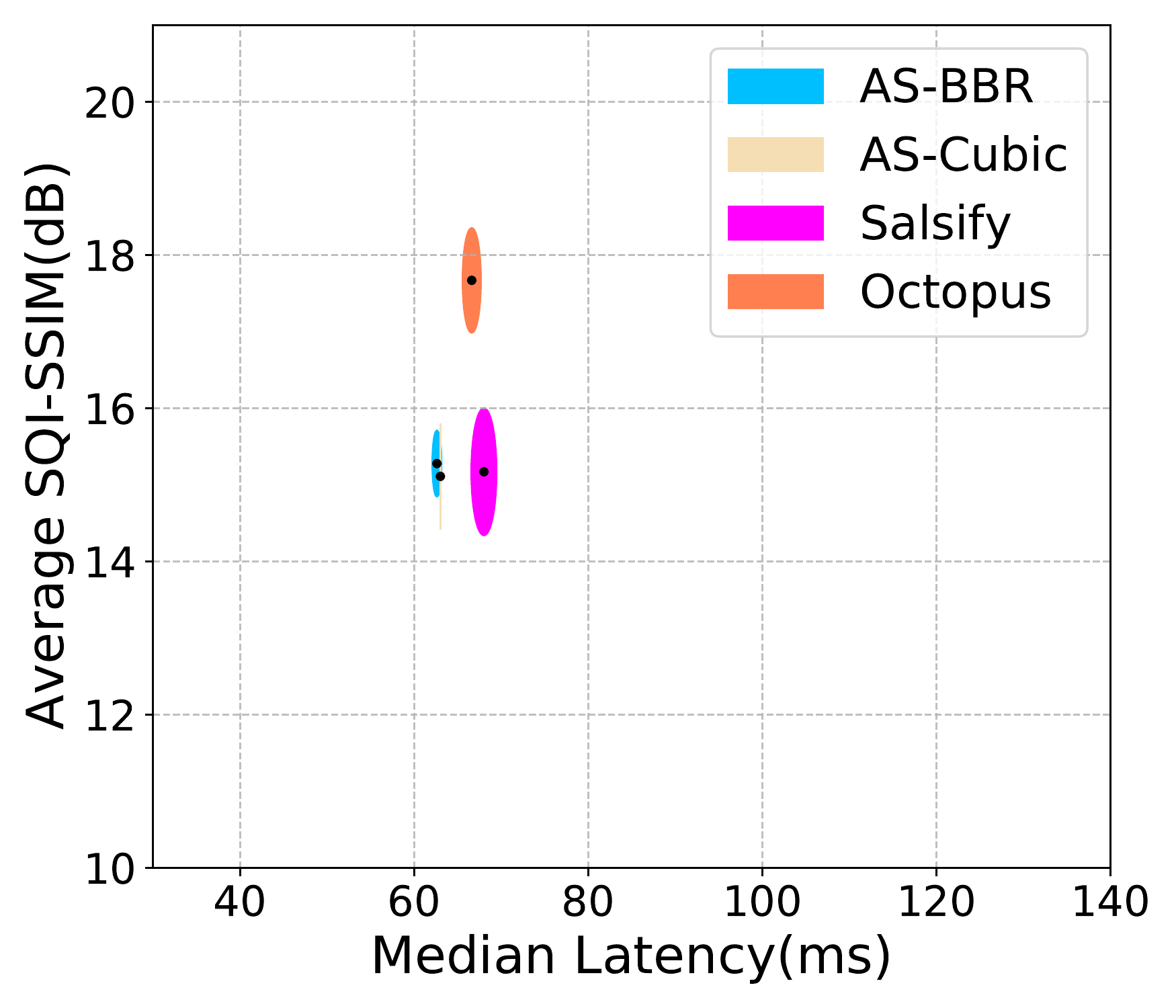}}\quad
    \subfigure[\normalsize{TMobile}]{\includegraphics[width=0.32\textwidth]{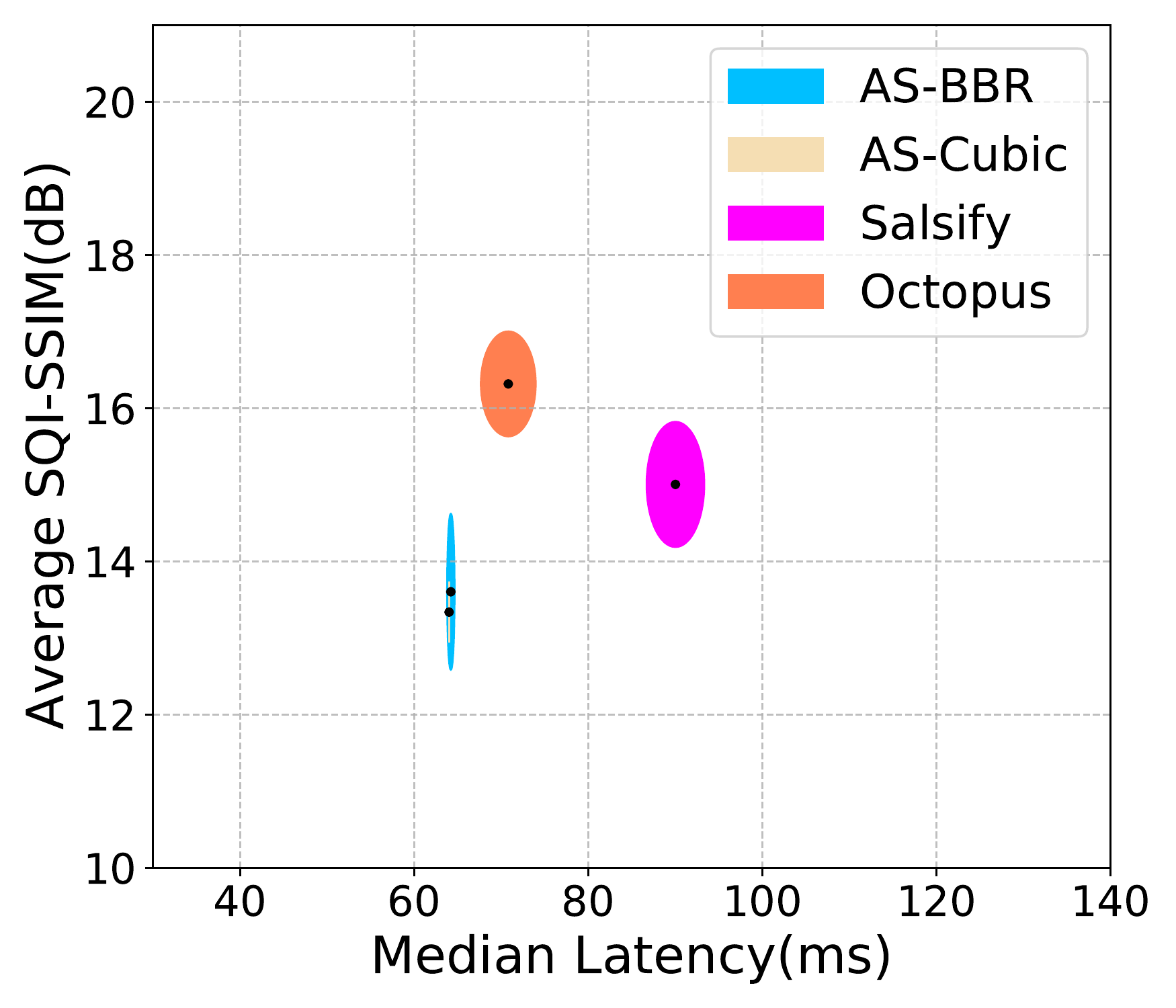}}\quad
    \subfigure[\normalsize{Verizon}]{\includegraphics[width=0.32\textwidth]{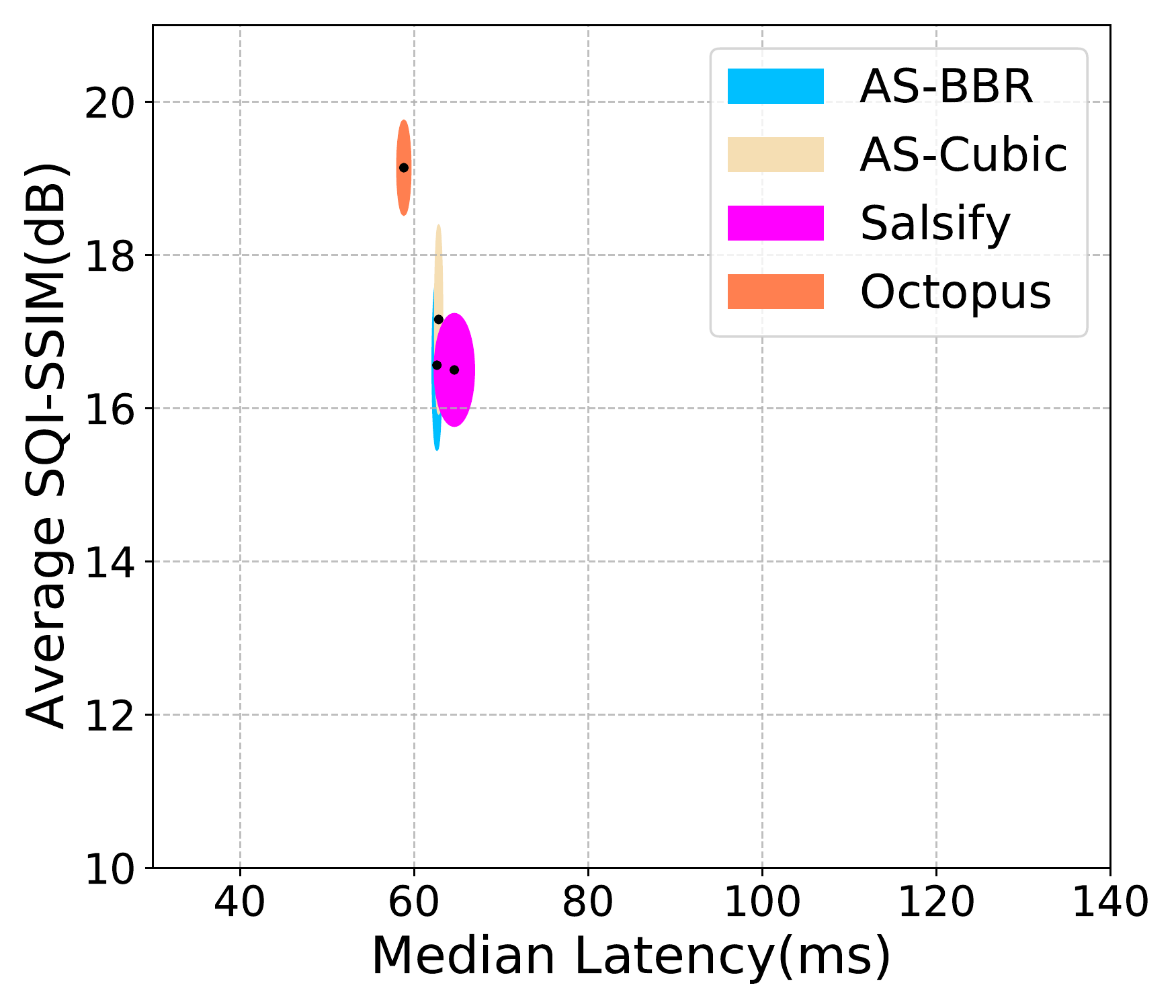}}
    \caption{The median latency v.s. video quality of Octopus and other baselines in different LTE download traces with RTT 60ms. The axes of the ellipse reflect the standard deviations in SQI-SSIM and median latency.}
    \label{fig:eval-usecase2-mid}
\end{figure*}

We provide several additional case study results in this section. Figure~\ref{fig:eval-usecase2-mid} shows the median latency and SQI-SSIM QoE metrics of Octopus and other baselines for real-time video with quality adaptation. All systems achieve similar low median latency, and Octopus has the highest video quality across all systems.

\begin{figure*}[h!]
    \centering
    \subfigure[\normalsize{ATT}]{\includegraphics[width=0.32\textwidth]{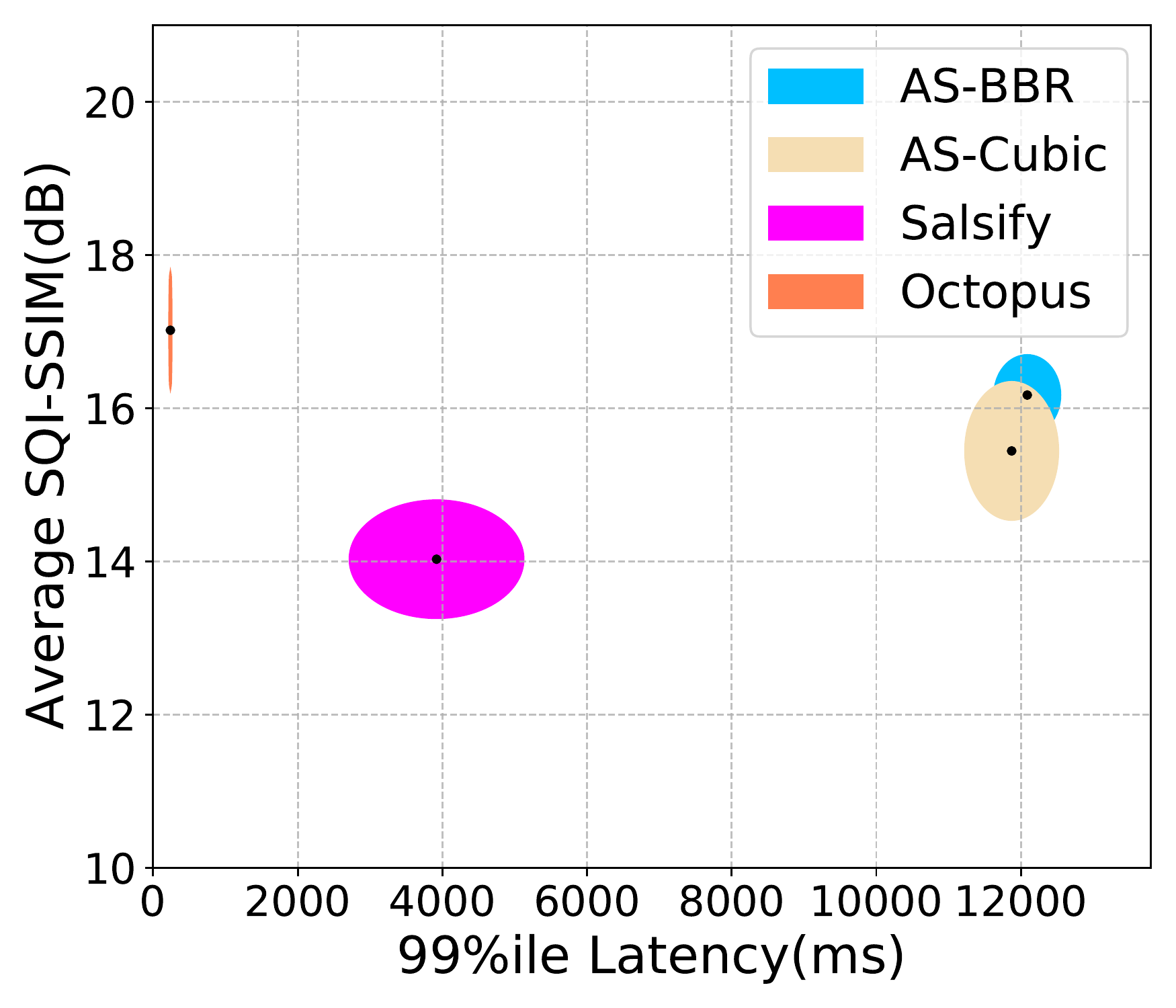}}\quad
    \subfigure[\normalsize{TMobile}]{\includegraphics[width=0.32\textwidth]{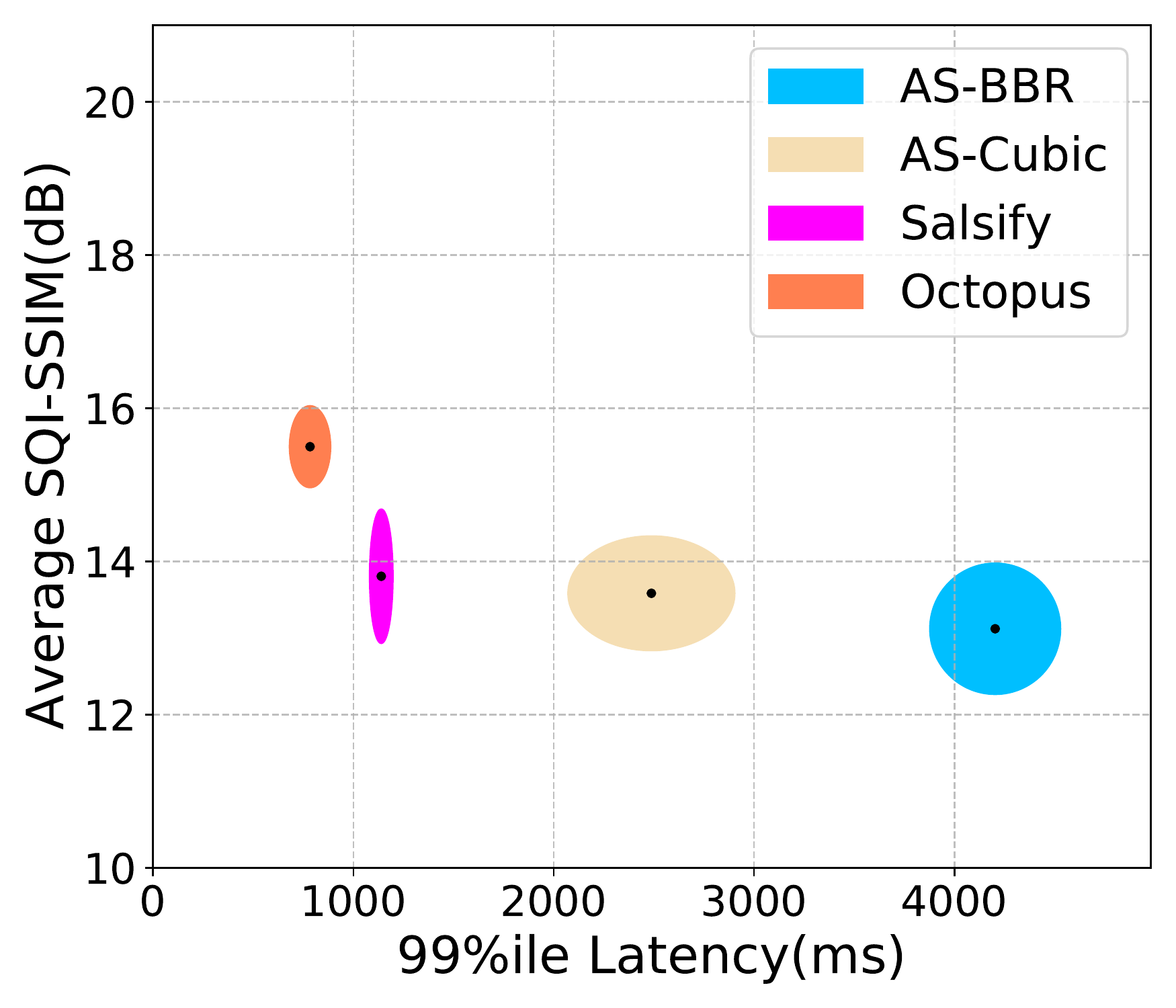}}\quad
    \subfigure[\normalsize{Verizon}]{\includegraphics[width=0.32\textwidth]{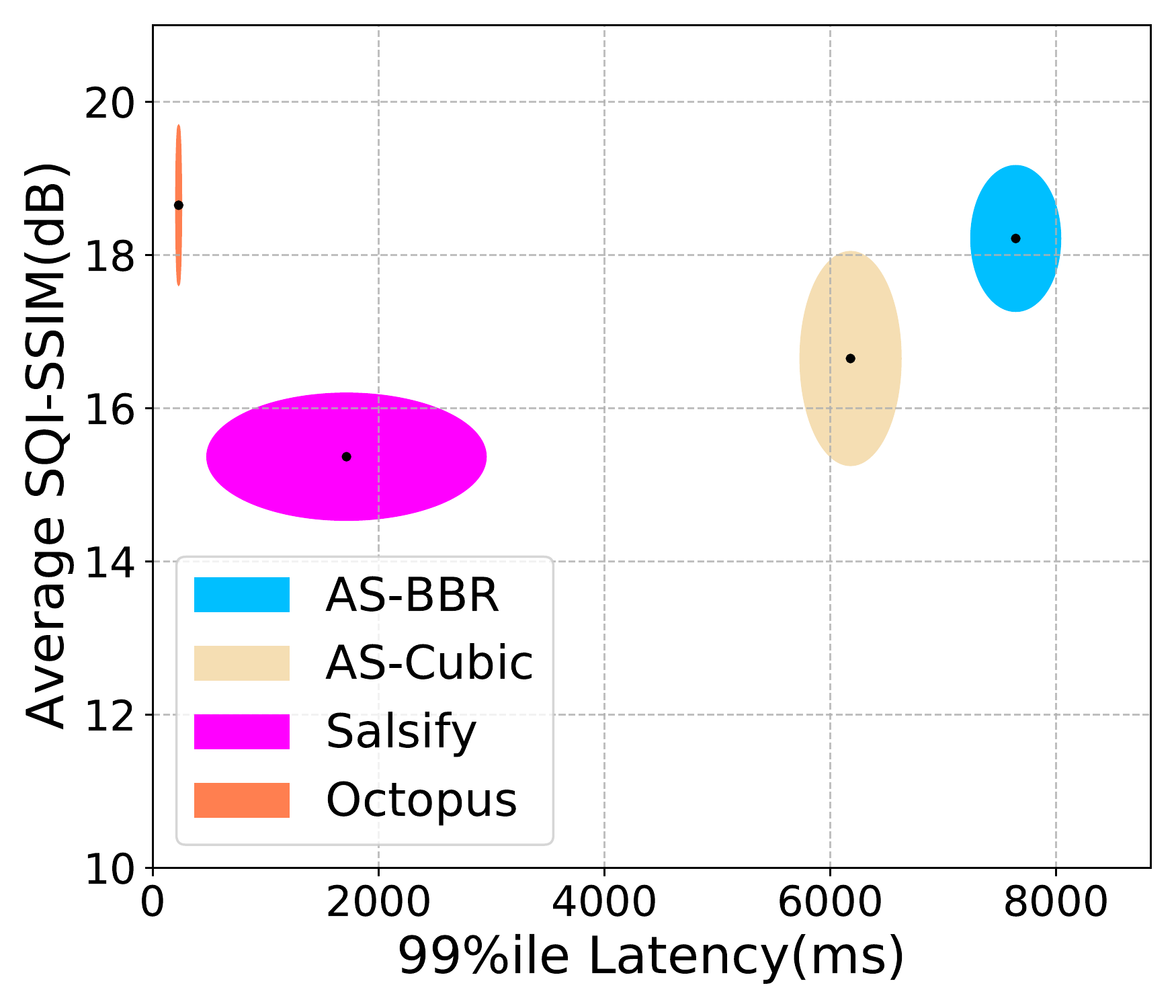}}
    \caption{The 99\%ile tail latency v.s. video quality of Octopus and other baselines in different LTE download traces with RTT 120ms.}
    \label{fig:eval-rtt120-tail}
\end{figure*}

\begin{figure*}[h!]
    \centering
    \subfigure[\normalsize{ATT}]{\includegraphics[width=0.32\textwidth]{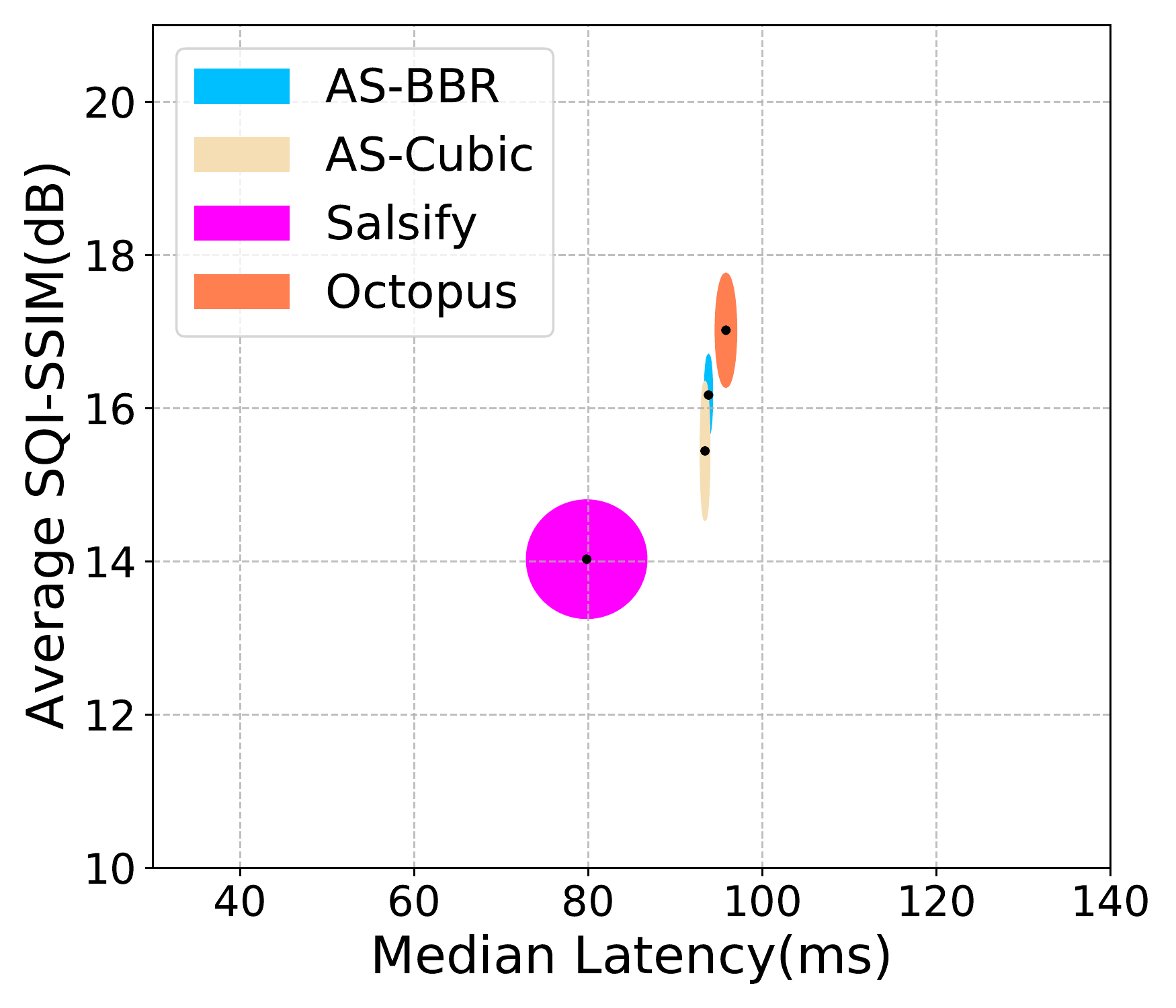}}\quad
    \subfigure[\normalsize{TMobile}]{\includegraphics[width=0.32\textwidth]{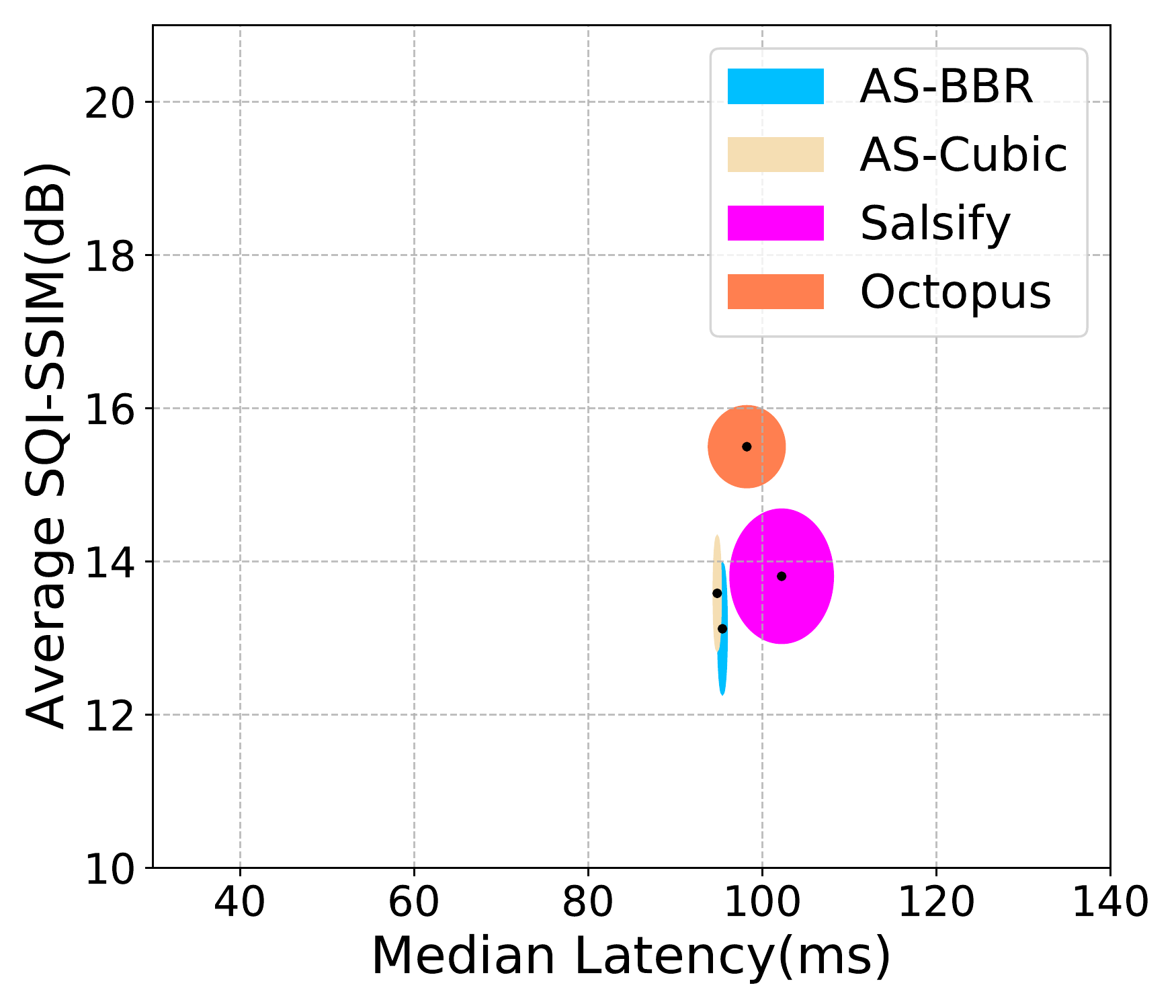}}\quad
    \subfigure[\normalsize{Verizon}]{\includegraphics[width=0.32\textwidth]{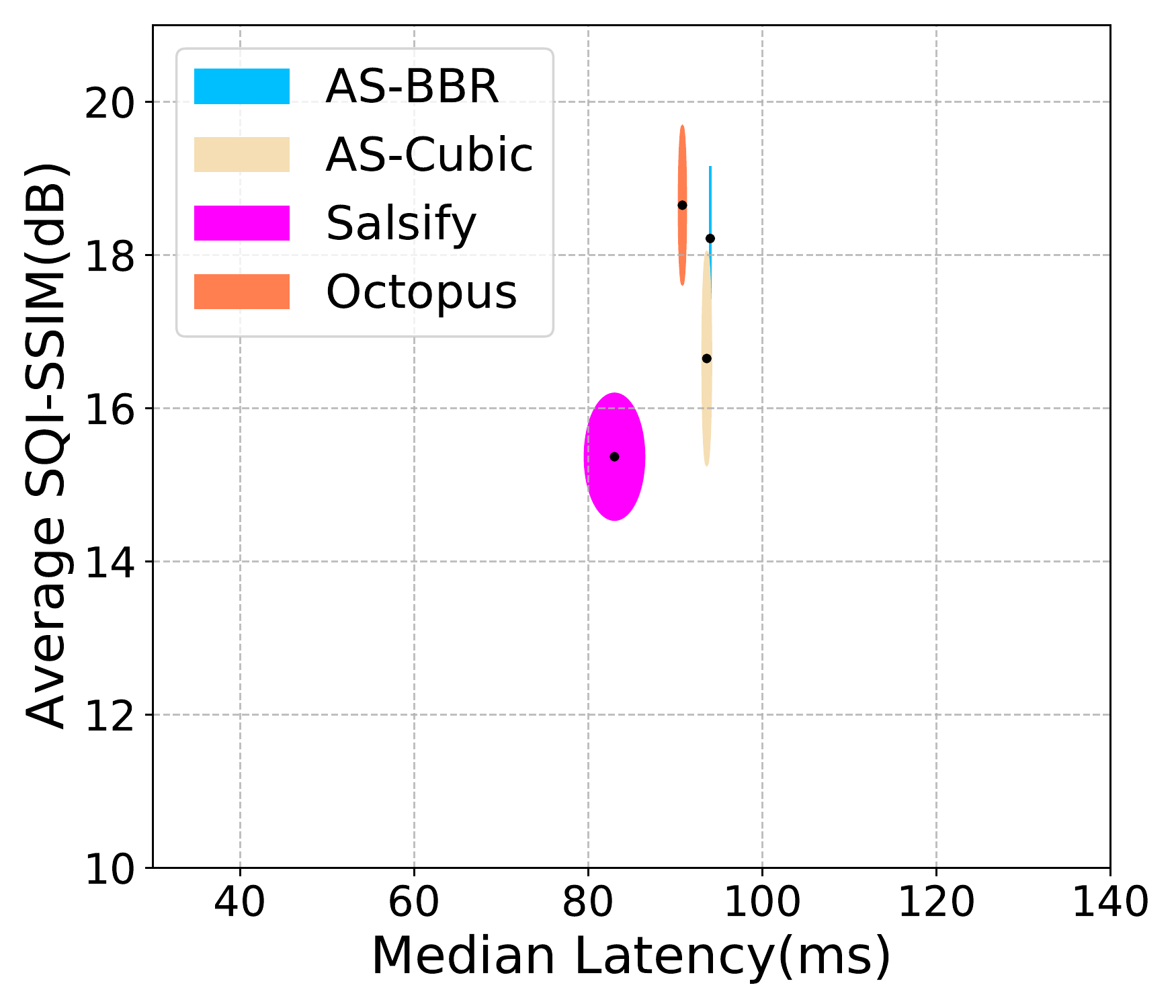}}
    \caption{The median latency v.s. video quality of Octopus and other baselines in different LTE download traces with RTT 120ms.}
    \label{fig:eval-rtt120-mid}
\end{figure*}

\begin{figure}[t!]
    \centering
    \subfigure[\footnotesize{5G trace1}]{\includegraphics[width=0.23\textwidth]{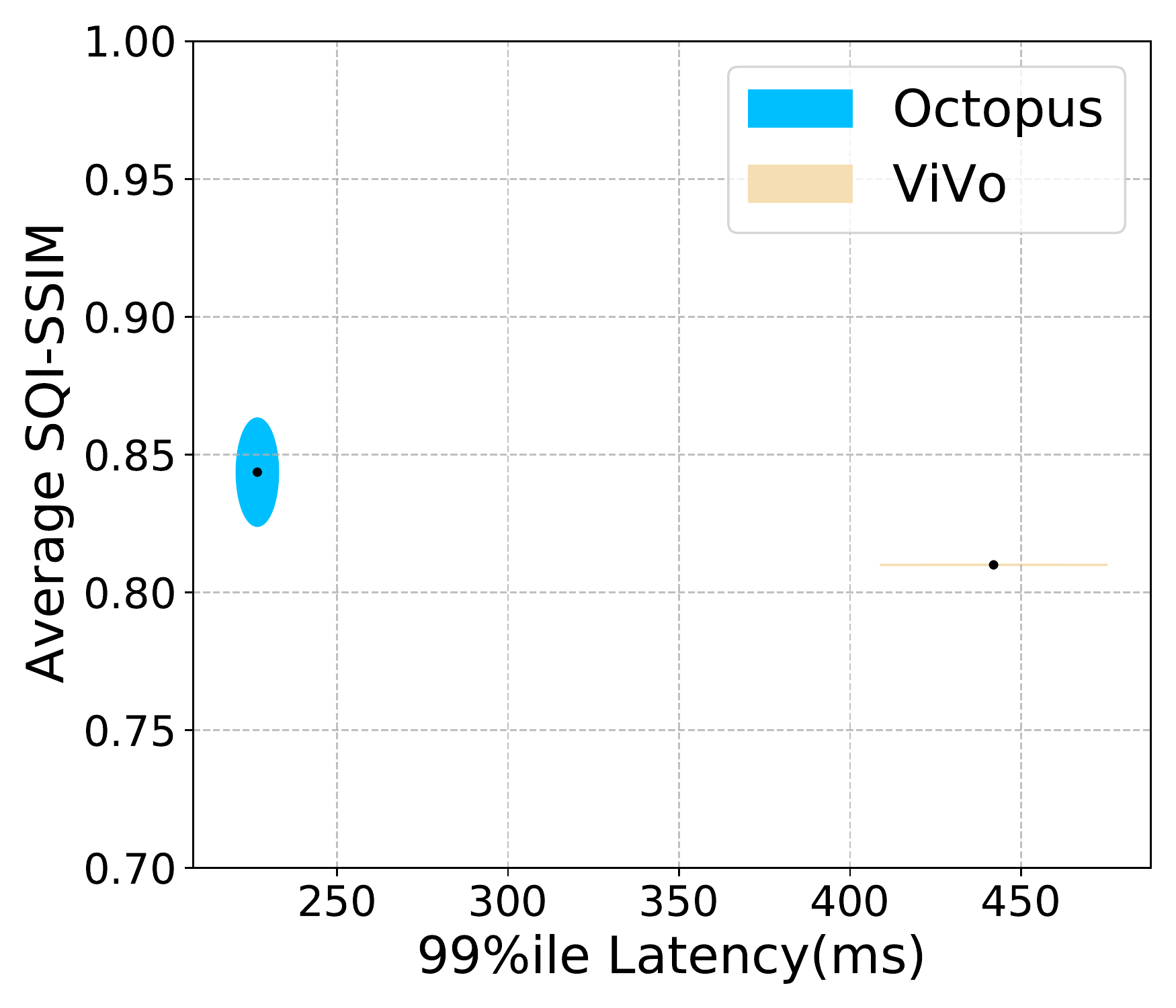}}
    \subfigure[\footnotesize{5G trace2}]{\includegraphics[width=0.23\textwidth]{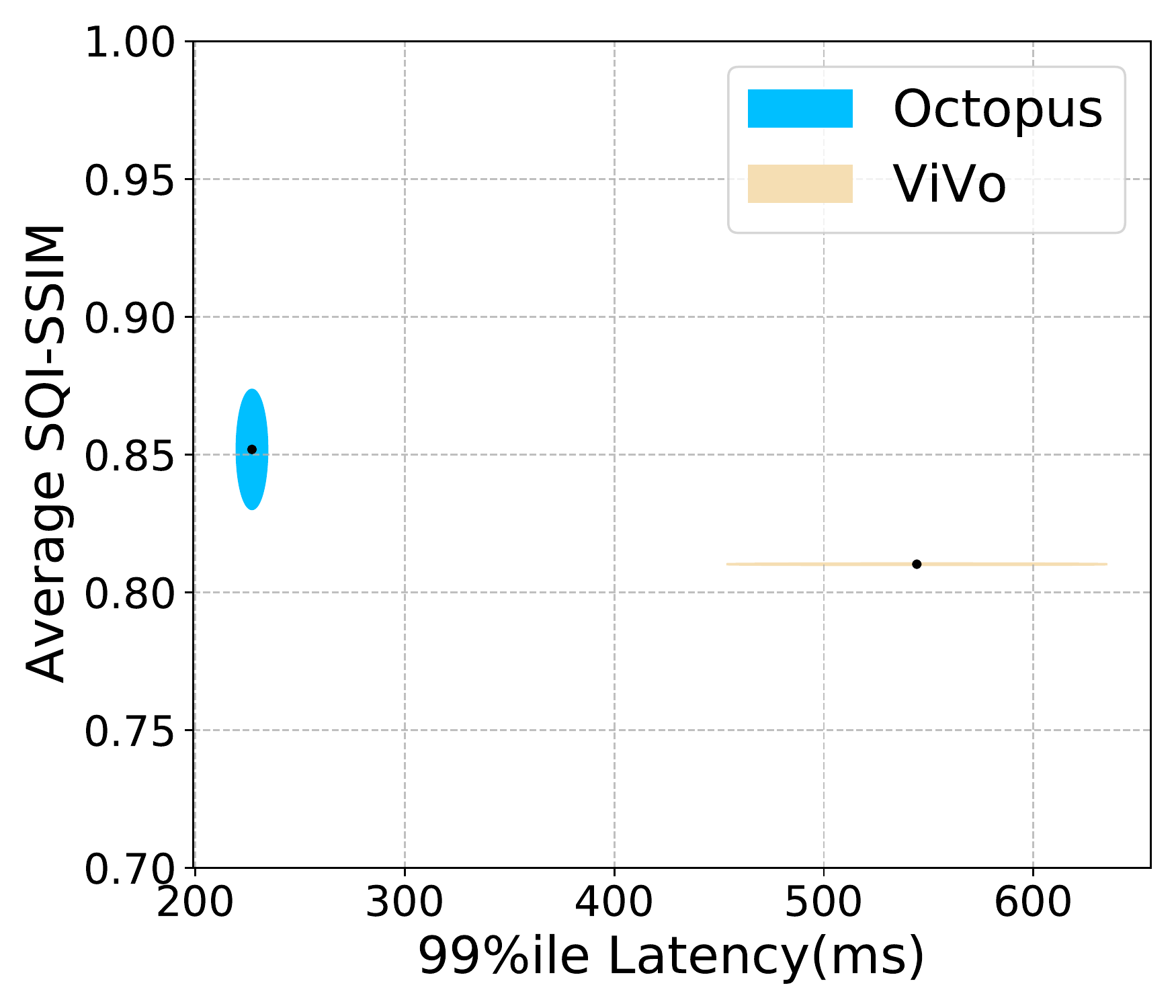}}
    \caption{The 99\% tail latency v.s. volumetric video quality of Octopus and ViVo in two 5G traces with RTT 120ms.}
    \vspace{-10pt}
    \label{fig:eval-usecase3-rtt120}
\end{figure}

Figure~\ref{fig:eval-rtt120-tail} and Figure~\ref{fig:eval-rtt120-mid} show the experiment results of the second case study with 120ms RTT. \oursys achieves similar performance improvement than other baselines. Compared with scenarios with 60ms RTT, all systems have lower video quality and slightly higher tail latency due to longer RTT and slower network feedback for congestion control. Figure~\ref{fig:eval-usecase3-rtt120} shows the experiment results of real-time volumetric video streaming with 120ms RTT. Octopus performs better than ViVo in both video quality and tail latency.

\section{Link Utilization on Legacy switches}
\label{appendix-link}
\begin{figure}[]
    \centering
    \includegraphics[width=0.45\textwidth]{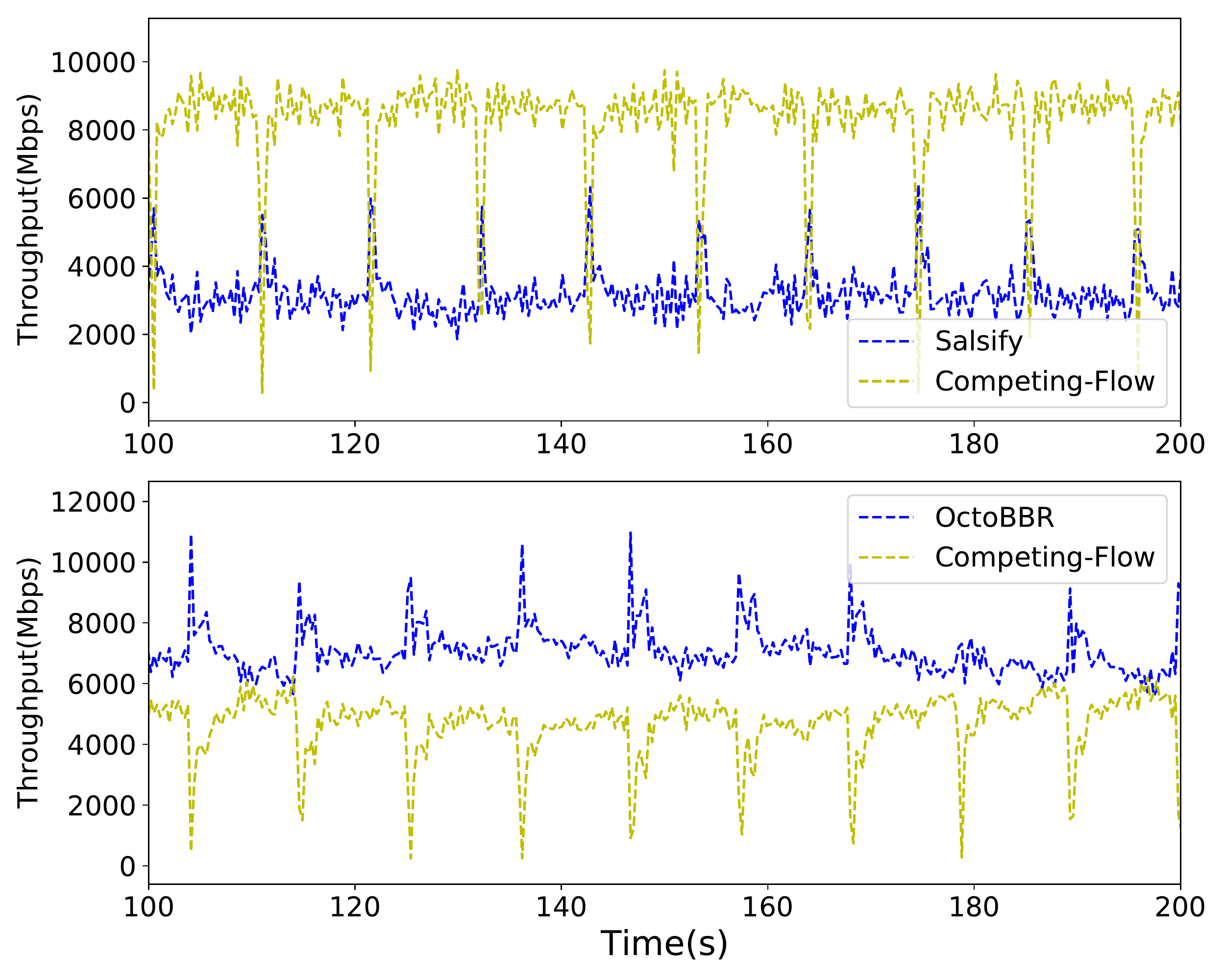}
    \caption{The throughput time series graph of OctoBBR and Salsify sharing the 12Mbps ethernet link with a competing flow.}
    \label{fig:bw-ethernet-compete}
\end{figure}

Figure~\ref{fig:bw-ethernet-compete} shows the link utilizations of OctoBBR and Salsify competing with a backlogged TCP flow on a legacy switch. In this case, OctoBBR earns a fair share of link capacity since its endhost congestion control is BBR. However, Salsify suffers from low link utilization due to its conservative behavior.

\section{Results with multiple bottlenecks}
\label{appendix-multiple}
\begin{figure}[]
    \centering
    \includegraphics[width=0.4\textwidth]{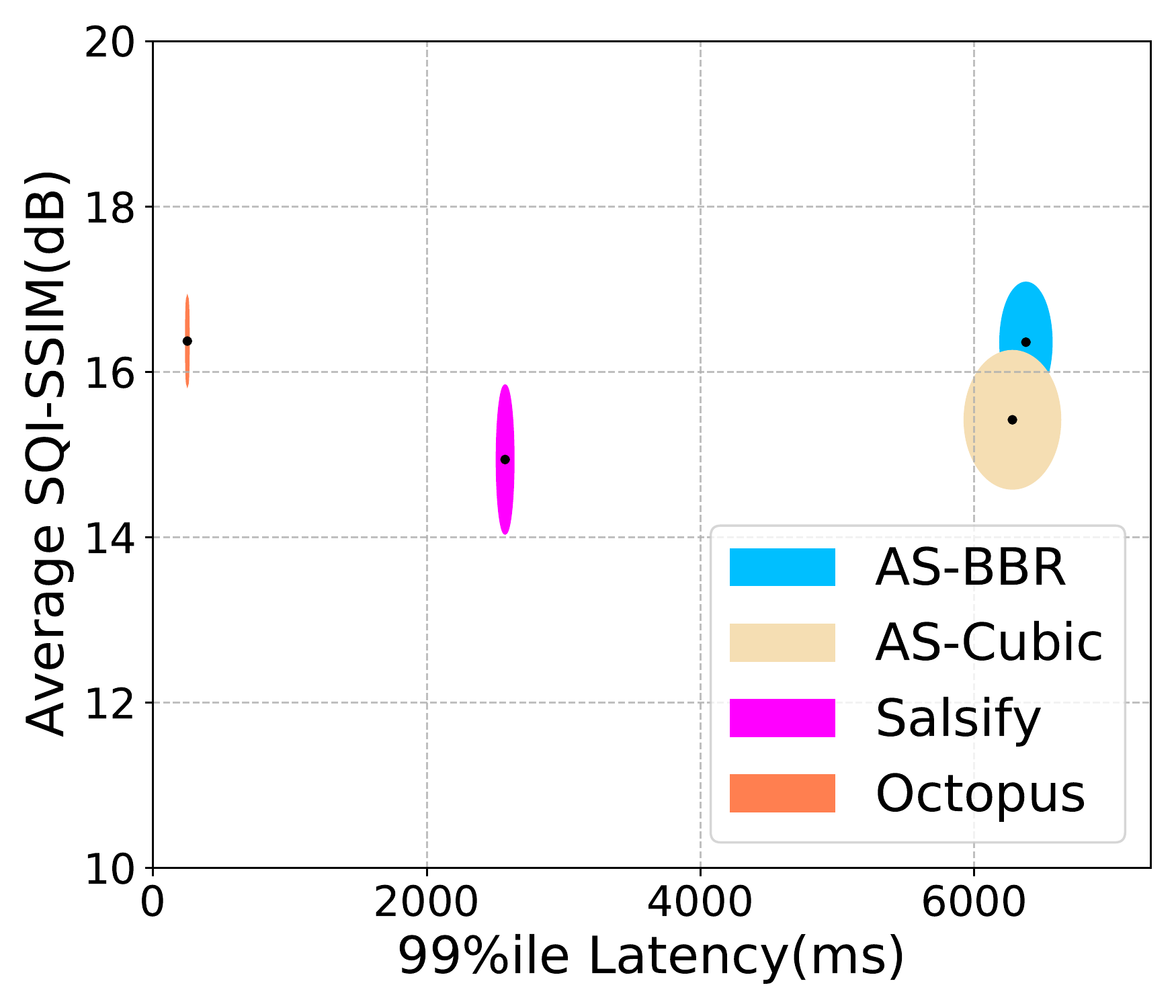}
    \caption{Real-time video with quality adaptation in scenarios with multiple bottlenecks.}
    \label{fig:multiple-bottleneck}
\end{figure}

In Figure~\ref{fig:multiple-bottleneck}, we evaluate Octopus and baselines from \S\ref{sec:casestudy2}, but in a scenario with multiple bottlenecks. The first bottleneck is at a legacy (non-\oursys) switch with a bandwidth of 12Mbps, and the second is a cellular (\oursys) link with bandwidths drawn from the Verizon downlink trace.  We see similar performance trends as those in \S\ref{sec:casestudy2}.

\end{appendices}

\end{document}